\documentclass[
  reprint,
  twocolumn,
  aps,
  prl,
  notitlepage,
  nobibnotes,
  longbibliography,
  superscriptaddress
]{revtex4-2}

\usepackage{amsmath,amssymb,amsthm,mathtools}
\usepackage{bm}

\usepackage{bbm}
\usepackage{dsfont}

\usepackage{graphicx}
\usepackage{tikz}
\usetikzlibrary{positioning}
\usepackage{float}
\usepackage{caption}
\usepackage{subcaption}

\usepackage{array}
\usepackage{enumitem}
\usepackage{xspace}
\usepackage{xparse}

\usepackage{xcolor}
\usepackage[
  colorlinks=true,
  linkcolor=blue,
  citecolor=blue,
  urlcolor=blue
]{hyperref}

\usepackage{subfiles}
\newif\ifwithSM
\withSMfalse

\DeclarePairedDelimiter{\paren}{(}{)}
\DeclarePairedDelimiter{\brkt}{[}{]}

\newcommand{\argmin}{\mathop{\mathrm{arg\,min}}}

\newcommand{\1}{\mathds{1}}

\newcommand{\B}{\mathcal B}
\newcommand{\D}{\mathcal D}

\renewcommand{\H}{\mathcal H}

\newcommand{\K}{\mathcal K}
\renewcommand{\L}{\mathcal L}

\renewcommand{\O}{\mathcal O}

\newcommand{\V}{\mathcal V}
\newcommand{\W}{\mathcal W}

\newcommand{\C}{\mathbb{C}}
\newcommand{\N}{\mathbb{N}}

\let\P\relax

\NewDocumentCommand{\P}{s o o g}{%
  \mathbb{P}%
  \IfValueT{#2}{_{#2}}%
  \IfValueT{#3}{^{#3}}%
  \IfValueTF{#4}{%
    \IfBooleanTF{#1}{\paren*{#4}}{\paren{#4}}%
  }{}%
}

\NewDocumentCommand{\PP}{s o o m m}{%
  \mathbb{P}%
  \IfValueT{#2}{_{#2}}%
  \IfValueT{#3}{^{#3}}%
  \IfBooleanTF{#1}
    {\brkt*{#4\middle\vert #5}}
    {\brkt{#4 \vert #5}}%
}

\NewDocumentCommand{\E}{s o o g}{%
  \mathbb{E}%
  \IfValueT{#2}{_{#2}}%
  \IfValueT{#3}{^{#3}}%
  \IfValueTF{#4}{%
    \IfBooleanTF{#1}{\brkt*{#4}}{\brkt{#4}}%
  }{}%
}

\NewDocumentCommand{\EE}{s o o m m}{%
  \mathbb{E}%
  \IfValueT{#2}{_{#2}}%
  \IfValueT{#3}{^{#3}}%
  \IfBooleanTF{#1}
    {\brkt*{#4 \middle\vert #5}}
    {\brkt{#4 \vert #5}}%
}

\NewDocumentCommand{\norm}{s o o m}{%
  \IfBooleanTF{#1}
    {\lVert #4 \rVert}
    {\left\lVert #4 \right\rVert}%
  \IfValueT{#2}{_{#2}}%
  \IfValueT{#3}{^{#3}}%
}

\NewDocumentCommand{\tr}{s o o g}{%
  \operatorname{tr}%
  \IfValueT{#2}{_{#2}}%
  \IfValueT{#3}{^{#3}}%
  \IfValueTF{#4}{%
    \IfBooleanTF{#1}{\paren*{#4}}{\paren{#4}}%
  }{}%
}

\NewDocumentCommand{\Tr}{s o o g}{%
  \operatorname{Tr}%
  \IfValueT{#2}{_{#2}}%
  \IfValueT{#3}{^{#3}}%
  \IfValueTF{#4}{%
    \IfBooleanTF{#1}{\paren*{#4}}{\paren{#4}}%
  }{}%
}

\NewDocumentCommand{\Trsq}{s o o g}{%
  \operatorname{Tr}%
  \IfValueT{#2}{_{#2}}%
  \IfValueT{#3}{^{#3}}%
  \IfValueTF{#4}{%
    \IfBooleanTF{#1}{\left[#4\right]}{[#4]}%
  }{}%
}

\let\ket\relax
\let\bra\relax

\NewDocumentCommand{\ket}{s m}{%
  \IfBooleanTF{#1}{\left| #2 \right\rangle}{|#2\rangle}%
}

\NewDocumentCommand{\bra}{s m}{%
  \IfBooleanTF{#1}{\left\langle #2 \right|}{\langle #2|}%
}

\NewDocumentCommand{\ketbra}{s m g}{%
  \IfBooleanTF{#1}
    {\left| #2 \middle\rangle\!\middle\langle \IfValueTF{#3}{#3}{#2} \right|}
    {|#2\rangle\langle \IfValueTF{#3}{#3}{#2}|}%
}
\newif\ifshowcomments
\showcommentstrue

\ifshowcomments
  \newcommand{\jpg}[1]{\textcolor{magenta}{\bf [JPG: #1]}}
\else
  \newcommand{\jpg}[1]{}
\fi

\newcommand{\ssp}{\nobreak\hspace{0.1em}}

\newcommand{\er}[1]{Eq.\ssp\eqref{#1}}
\newcommand{\ers}[2]{Eqs.\ssp(\ref{#1}--\ref{#2})}

\newtheorem{theorem}{Theorem}
\newcommand{\fb}{{\rm fb}}

\begin{document}
\title{
    Generating quantum ensembles via reverse-time quantum diffusions
    }

\author{Ma\"el Bompais} \email{mael.bompais@nottingham.ac.uk}
\affiliation{School of Mathematical Sciences, University of Nottingham, University Park, Nottingham, NG7 2RD, United Kingdom}
\affiliation{Centre for the Mathematics and Theoretical Physics of Quantum Non-Equilibrium Systems, University of Nottingham, Nottingham, NG7 2RD, UK}

\author{M\u{a}d\u{a}lin Gu\c{t}\u{a}}
\affiliation{School of Mathematical Sciences, University of Nottingham, University Park, Nottingham, NG7 2RD, United Kingdom}
\affiliation{Centre for the Mathematics and Theoretical Physics of Quantum Non-Equilibrium Systems, University of Nottingham, Nottingham, NG7 2RD, UK}

\author{Juan P. Garrahan}
\affiliation{School of Physics and Astronomy, University of Nottingham, University Park, Nottingham, NG7 2RD, United Kingdom}
\affiliation{Centre for the Mathematics and Theoretical Physics of Quantum Non-Equilibrium Systems, University of Nottingham, Nottingham, NG7 2RD, UK}

\begin{abstract}
We establish a reverse-time denoising theory for quantum diffusions of continuously measured quantum systems. Starting from the stochastic Schr\"{o}dinger equation of a forward noising dynamics, we derive the exact reverse-time dynamics for quantum trajectories, whose law coincides with the time-reversal of the original process. 
We prove that the denoising dynamics is a physically admissible quantum diffusion, with the same measurement-induced noise but a state-dependent feedback Hamiltonian, a direct analogue of the ``score function'' of generative classical diffusion models. This provides a principled framework for converting samples of a simple distribution into those of a more complex ensemble of quantum states. We show how the denoising dynamics can be directly learnt from forward trajectory data, and how to exploit purification to initialise the denoising process.
\end{abstract}

\maketitle

\noindent
{\bf \em Introduction.}
Generative modelling aims at learning a mechanism capable of producing new samples from a target distribution \cite{bond2021deep}. In recent years, diffusion models have emerged as a powerful framework for classical data generation \cite{sohl-dickstein2015deep,ho2020denoising,song2021score-based}. Their principle is simple: data samples are progressively corrupted by a stochastic noising dynamics, and new samples are generated by learning a suitable reverse-time denoising dynamics that transports noise back to the data distribution. This approach is grounded in the classical theory of reverse-time diffusions \cite{anderson1982reverse-time}, which provides the mathematical foundation underlying the framework.

Recent works have begun to explore quantum analogues of diffusion-based generative modelling. One line of research considers {\em deterministic} dynamics in which the forward noising process is implemented through a sequence of quantum channels acting on the state, and the reverse denoising dynamics is realised by learned quantum operations \cite{chen2024quantum,kwun2025mixed-state,parigi2025quantum-noise-driven,nasu2025quantum}. In parallel, others have investigated {\em stochastic} formulations closer
to the classical diffusion framework, where the evolution of the quantum
state is driven by random circuits, measurements, or external classical
noise \cite{zhang2024generative,liu2025measurement-based}.
Related work has also considered record-conditioned reversal of monitored
quantum trajectories for Pauli noise \cite{gabbassov2025exact}.

\begin{figure}[!t]
\includegraphics[width=0.8\columnwidth]{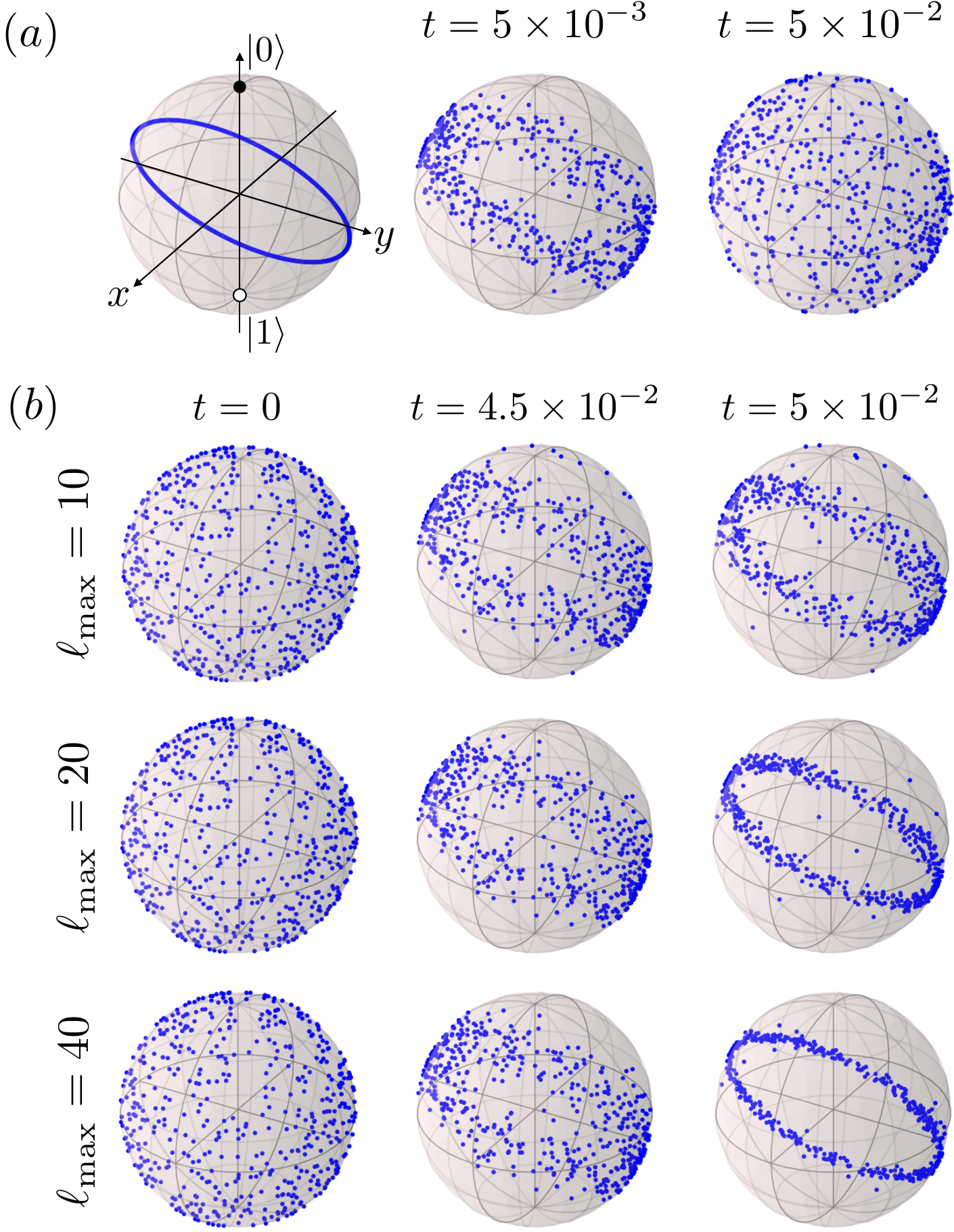}
  \caption{
    (a) Forward (noising) dynamics for the 2-level example for samples from an initial measure over pure states $\mu_0(\psi)$ corresponding to a great circle on the Bloch sphere. Under the noising dynamics the samples evolve towards a uniform $\mu_{\rm inv}(\psi) = 1/4\pi$. 
    (b) Backward denoising dynamics starting from uniformly distributed samples, for various levels $\ell_{\rm max}$ of approximation of the score, see \cite{see-supplemental} for details. At long times the initial $\mu_0(\psi)$ is reconstructed, thus transforming noisy initial pure states to samples of the distribution of interest. 
    \label{fig:bloch_spheres}
    }    
\end{figure}

While these works demonstrate that diffusion-based ideas can be adapted to
the quantum setting, important limitations remain. Existing generative
constructions are generally restricted to specially designed dynamics, and
theoretical guarantees for the correctness of the full stochastic process
are incomplete or absent. In particular, a general reverse-time
construction for continuously monitored quantum systems, valid for
arbitrary Hamiltonian and coupling operators and directly realisable as a
quantum stochastic trajectory, is still lacking.

In this work, we present a full time-reversal of quantum trajectories driven by diffusive bosonic noise processes, for arbitrary Hamiltonian and coupling operators. Our approach establishes an equality in law at the level of stochastic processes, rather than only at the level of marginal state distributions. This allows the forward quantum trajectories to be used directly for training the denoising model. More broadly, the resulting framework addresses all the issues raised for the previous works. Our stochastic formulation also provides several practical advantages for state preparation, learning, and monitoring of the generative process. In particular, the reverse-time denoising dynamics takes the form of a feedback-controlled quantum diffusion whose control Hamiltonian can be written explicitly and depends, as in the classical case, on a suitable score function (see Theorem \ref{thm:main}). Furthermore, we discuss how to define training depending on the information available, ranging from full access to forward trajectories to situations where the initial states are unknown and must be estimated. We illustrate our general approach with a simple example of sampling a non-trivial distribution of pure states of a qubit.

\bigskip
\noindent
{\bf \em Quantum diffusion as forward noising dynamics.} We consider a finite-dimensional Hilbert space $\H$ and denote by $\B(\H)$ the set of linear operators on $\H$. We focus on pure states $\psi = \ket{\psi}\bra{\psi}$. The {\em forward}, or {\em noising}, dynamics is given at the trajectory level by the stochastic Schr\"{o}dinger equation (SSE) 
\cite{belavkin1990a-stochastic,gardiner2004quantum,breuer2002the-theory},
\begin{equation}
  \label{eq:SSE}
  d \psi_t 
  = \L(\psi_t)\, dt 
    + \sum_{m=1}^M \K_m(\psi_t)\, dW_t^m \, .  
\end{equation}
Here $(W_t^m)_{m=1}^M$ are independent standard Wiener processes, the drift term is given by the Lindblad generator
\begin{align}
  \label{eq:Lindblad}
  \L(\cdot)
  & \coloneqq 
  -i[H,(\cdot)] + \D(\cdot)
    \, ,
\end{align}
with dissipator
\begin{align}
  \label{eq:Dissipator}
  \D(\cdot)  
  & \coloneqq 
  \sum_{m=1}^M 
    \left(
      L_m (\cdot) L_m^\dag
      - \frac12 \{ L_m^\dag L_m, (\cdot) \}
    \right)
    \, ,
\end{align}
and the diffusion coefficients are given by
\begin{align}
  \label{eq:K}
  \K_m(\cdot)
  & \coloneqq 
  L_m (\cdot) + (\cdot) L_m^\dag
  - \Trsq{L_m (\cdot) + (\cdot) L_m^\dag} \, (\cdot)
  \, .
\end{align}
In \ers{eq:Lindblad}{eq:K} $H$ is a Hamiltonian and $L_m \in \B(\H)$ are coupling operators describing the coupling to $M$ environment bosonic channels. The SSE \eqref{eq:SSE} preserves purity and represents the  system evolution conditional on the trajectory of homodyne measurements in environment channels. This can be seen as an unravelling of the Lindblad semigroup generated by $\L$, so that the average state, $\rho_t := \E{\psi_t}$, evolves according to the Gorini-Kossakowski-Sudarshan-Lindblad master equation \cite{lindblad1976on-the-generators,gorini1976completely},
\begin{equation}
  \label{eq:masterequation}
  \frac{d}{dt}\rho_t = \L(\rho_t) 
  \, .
\end{equation}

For the generation task, we will assume that the master evolution is \emph{irreducible},
which means that there exists a unique stationary 
state $\rho_{\rm inv}$ such that $\mathcal{L}(\rho_{\rm inv})=0$. 
This ensures, together with a purification condition \cite{benoist2021invariant}, that the stochastic evolution has a unique invariant distribution $\mu_{\mathrm{inv}}$ over pure states and  
is exponentially mixing: if $\mu_t$ denotes the law of $\psi_t$, then one has exponential convergence
\begin{equation}
\label{eq:exp.mixing}
  W_1(\mu_t,\mu_{\mathrm{inv}})
  \le C e^{-\lambda t},
\end{equation}
for some constants $C>0$ and $\lambda>0$, 
where $W_1$ denotes the $1$-Wasserstein distance between probability distributions.

In analogy to classical diffusion models, the aim of the forward (noising) process is to convert a structured distribution $\mu_0$ (what one wishes to sample) into a structureless $\mu_{\rm{inv}}$ (which is easy to sample). 
The law at time $t$ is given by the pushforward of the initial distribution $\mu_0$ by the stochastic flow $F$ associated with the SSE, $\mu_t = F_t\mu_0.$
As $\mu_t$ converges exponentially fast to $\mu_{\mathrm{inv}}$, in practice the forward evolution is stopped at a sufficiently large finite time $T$, for which $\mu_T$ is close to $\mu_{\mathrm{inv}}$.

\bigskip
\noindent
{\bf \em Reverse-time denoising dynamics.} 
Our goal is to construct a {\em reverse-time} or {\em denoising} stochastic dynamics
which maps $\mu_T \approx \mu_{\rm inv}$ back to $\mu_0$ at the level of trajectories. With this we can imagine sampling a pure state from $\mu_{T}$ and, by running this backward dynamics, recover a sample from $\mu_0$. 

The problem can be formulated as follows. If $(\psi_t)_{t\in[0,T]}$ is the forward process with marginal law $\mu_t$, we seek another process $(\tilde \psi_t)_{t\in[0,T]}$ such that
\begin{equation}
  \label{eq:def_reverse_in_law}
  ( \tilde \psi_t)_{t\in[0,T]}
  \ \stackrel{\mathrm{law}}{=}\ 
  (\psi_{T-t})_{t\in[0,T]} 
  \, , 
\end{equation}
and in particular, if $\tilde \psi_0 \sim \mu_T$ then $\tilde \psi_T \sim \mu_0$. Thus, the reverse process  transforms samples drawn from the ``noise'' distribution $\mu_T$ into samples distributed according to the ``data'' distribution $\mu_0$, while reproducing the statistics of the forward diffusion trajectories but with time reversed. 

Beyond equality in law \eqref{eq:def_reverse_in_law}, we impose that the reverse process be \emph{physically admissible}. More precisely, we require that $(\tilde\psi_t)$ be described by an SSE similar to \er{eq:SSE}, but potentially with time-dependent, and in general state-dependent, drift and diffusion coefficients, 
\begin{equation}
  d \tilde \psi_t
  =
  \tilde{\L}_t(\tilde \psi_t)\, dt
  +
  \sum_{m=1}^M
  \tilde{\K}_{m,t}(\tilde \psi_t)\, d\tilde W_t^m
  \, .
\end{equation}

We now provide our main result (see \cite{see-supplemental} for the full details of the proof).

\begin{theorem}\label{thm:main}
Let $(\psi_t)_{t\in[0,T]}$ satisfy the stochastic Schrödinger equation
\eqref{eq:SSE}, and let $\mu_t$ denote its marginal law. Then there exists a physically admissible quantum diffusion
$(\tilde\psi_t)_{t\in[0,T]}$ which is the time reversal of $(\psi_t)_{t\in[0,T]}$ in the sense of equation \eqref{eq:def_reverse_in_law}.
The reverse-time evolution retains the stochastic term and the dissipative part of the drift of the forward evolution, but differs through a feedback control Hamiltonian. Specifically, its dynamics is governed by
\begin{align}
\label{eq:reverse_SSE}
d\tilde\psi_t
=
\left(
-i[ H+  H^{\rm fb}_t(\tilde\psi_t),\tilde\psi_t]
+\mathcal D(\tilde\psi_t)
\right)dt
\nonumber 
\\
+
\sum_{m=1}^M
\mathcal K_m(\tilde\psi_t)\,d\tilde W_t^m,
\end{align}
where $H_t^\fb$ is a state- and
time-dependent Hamiltonian that can be chosen as  
\begin{equation}
\label{eq:feedback_hamiltonian}
 H_t^\fb(\psi)
=
- 2H + i[X_t(\psi),\psi],
\end{equation}
with
\begin{equation}
X_t(\psi)
=
-2\mathcal D(\psi)
+\operatorname{div}D(\psi)
+
D(\psi)\nabla_\psi\log\mu_{T-t}(\psi),
\label{eq:X}
\end{equation}
where $D$ is the diffusion tensor \cite{carollo2021large}, with components
\begin{equation}
\label{eq:diffusion_tensor}
D_{ij,hk}(\psi)
=
\sum_{m=1}^M
\bigl(\mathcal K_m(\psi)\bigr)_{ij}
\bigl(\mathcal K_m(\psi)\bigr)_{hk},
\end{equation}
and its partial divergence $\operatorname{div}D$ is defined component-wise by
\begin{equation}
\label{eq:partial_divergence}
(\operatorname{div}D)_{ij}(\psi)
=
\sum_{h,k}
\frac{\partial}{\partial\psi_{hk}}
D_{ij,hk}(\psi),
\end{equation}
where $\psi_{hk}$ denotes the $(h,k)$ matrix entry of $\psi$.
\end{theorem}

Since $H_t^{\mathrm{fb}}(\psi)$ is self-adjoint by construction, the reverse drift is again of Lindblad form, with a dissipator consistent with the corresponding noise operators, and the reverse SSE preserves the set of pure states.
It is important to note that the drift in \er{eq:reverse_SSE} is not the continuous-time version \cite{kwon2022reversing} of the Petz map associated to \er{eq:Lindblad}, or one of its  generalisations \cite{petz1986sufficient,parzygnat2023from,bai2025quantum}: the reversal is at the level of stochastic trajectories and ensembles of states, while the Petz map only reverses the average state. This stronger property is what makes training on forward data possible, as we show later.

\bigskip
\noindent
{\bf \em Initial-state preparation for denoising: information and robustness.
} The reverse process must be initialised with $\tilde\psi_0 \sim \mu_T$. However, direct preparation of $\mu_T$ may be challenging in practice, particularly in an experimental setting. 
If we choose the noising dynamics \eqref{eq:SSE} to be irreducible and to satisfy the purification condition \cite{maassen2006purification,amini2021on-asymptotic,benoist2021invariant,bompais2026the-rate}, a natural solution is to start from any state, run the forward dynamics for a time $T$ large enough so that the law at time $T$ becomes independent of this arbitrary initial condition, and in that case $\psi_T$ becomes an independent sample of $\mu_T \approx \mu_{\mathrm{inv}}$. Moreover, purification implies stability of the associated quantum filter, allowing the realised state to be inferred asymptotically from the measurement record alone \cite{amini2021on-asymptotic,bompais2026the-rate}. Thus, even if the arbitrary initial state is mixed, the state at time $T$ becomes approximately pure and known from the observed emissions.

This forward evolution prior to the denoising process therefore plays two roles: (a) it prepares a sample from $\mu_T \approx \mu_{\mathrm{inv}}$, and (b) through continuous measurement and purification, it provides precise knowledge of the pure state $\psi_T$ (which will be used as $\tilde{\psi}_0$ for denoising). The knowledge from (b) is crucial because the reverse Hamiltonian $ H_t^\fb(\tilde\psi)$ depends explicitly on the instantaneous state. This state has to be kept track of while denoising, since the backward control \eqref{eq:feedback_hamiltonian} depends on it. 

Note that the reverse evolution remains stochastic. Therefore the measurement-induced quantum diffusion continues to extract information about the system during the controlled dynamics. As a consequence, the denoising procedure benefits from continuous state estimation, which provides intrinsic robustness with respect to
imperfections in the control implementation.

\bigskip
\noindent
{\bf \em Learning the denoising dynamics from data.} While the dynamics \ers{eq:reverse_SSE}{eq:partial_divergence} is the exact solution to the denoising problem, it requires the full knowledge of the distribution of interest $\mu_0$, and its evolved version $\mu_t$. However, and just like in the classical generative case \cite{ho2020denoising,song2021score-based}, in most practical applications the control Hamiltonian $H^\fb$ will need to be approximated from samples
(i.e.\ the denoising model has to be ``trained''). The key to do this is the equality in law \eqref{eq:def_reverse_in_law}, which allows us to use forward trajectories, reversed in time, as sample trajectories of the backward process. This makes it possible to learn the backward dynamics from forward data. Depending on the amount of information available about the forward trajectories, we distinguish two levels of training.

\smallskip
\noindent
{\em Level 1: Training with access to full forward trajectories.} The ideal situation is when the initial state of each forward trajectory is known and the full trajectory can be reconstructed from the measurement signal.

Suppose we have $R$ training trajectories, $\{\psi_t^{(r)}\}_{r=1}^R$. From these we construct backward trajectories by time reversal, $\tilde\psi_t^{(r)} = \psi_{T-t}^{(r)}$. We parametrise the backward feedback Hamiltonian as
\begin{equation}
  \label{eq:Htheta}
  {H}_t^{\fb,\theta}  (\psi)
  \coloneqq
  \sum_{j=1}^J 
    \theta_t^{(j)}(\psi) \, H_j ,
\end{equation}
where $\{H_j\}_{j=1}^J$ is a fixed set of Hamiltonians, and $\theta_t \coloneqq  (\theta_t^{(1)}, \ldots, \theta_t^{(J)} ) \in \Theta$ 
are time-dependent functions to be learned from a chosen class of functions $\Theta$, such as neural networks or symmetry-constrained parametrisations.

Learning the Hamiltonian amounts to fitting the drift of the backward equation with the parametrisation \eqref{eq:Htheta}. Discretising time on a uniform grid $t_k = k\,\Delta t$, we define
\begin{equation}
  \label{eq:discretisation}
  \tilde{\psi}_k^{(r)} \coloneqq \tilde{\psi}_{t_k}^{(r)}
  \, ,
  \;\;\;
  R_k^{(r)} 
  \coloneqq
  \frac{\tilde{\psi}_{k+1}^{(r)} - \tilde{\psi}_{k}^{(r)}}{\Delta t}
  -
  \D(\tilde{\psi}_k^{(r)})
  \, + i[H,\tilde \psi_k^{(r)}] .
\end{equation}
The estimated parameters $\hat{\theta}$ are obtained by solving the regression problem
\begin{equation}
  \hat \theta
  =
  \argmin_{\theta\in\Theta}
    \sum_{r,k}
      \Big\|
        R_k^{(r)}
        + i 
        \sum_j
        \theta_{t_k}^{(j)}(\tilde{\psi}_k^{(r)})
        \, 
        [H_j , \tilde{\psi}_k^{(r)}]
      \Big\|_2^2 
  \, .
\end{equation}

\begin{figure}[t]
  \includegraphics[width=0.9\columnwidth]{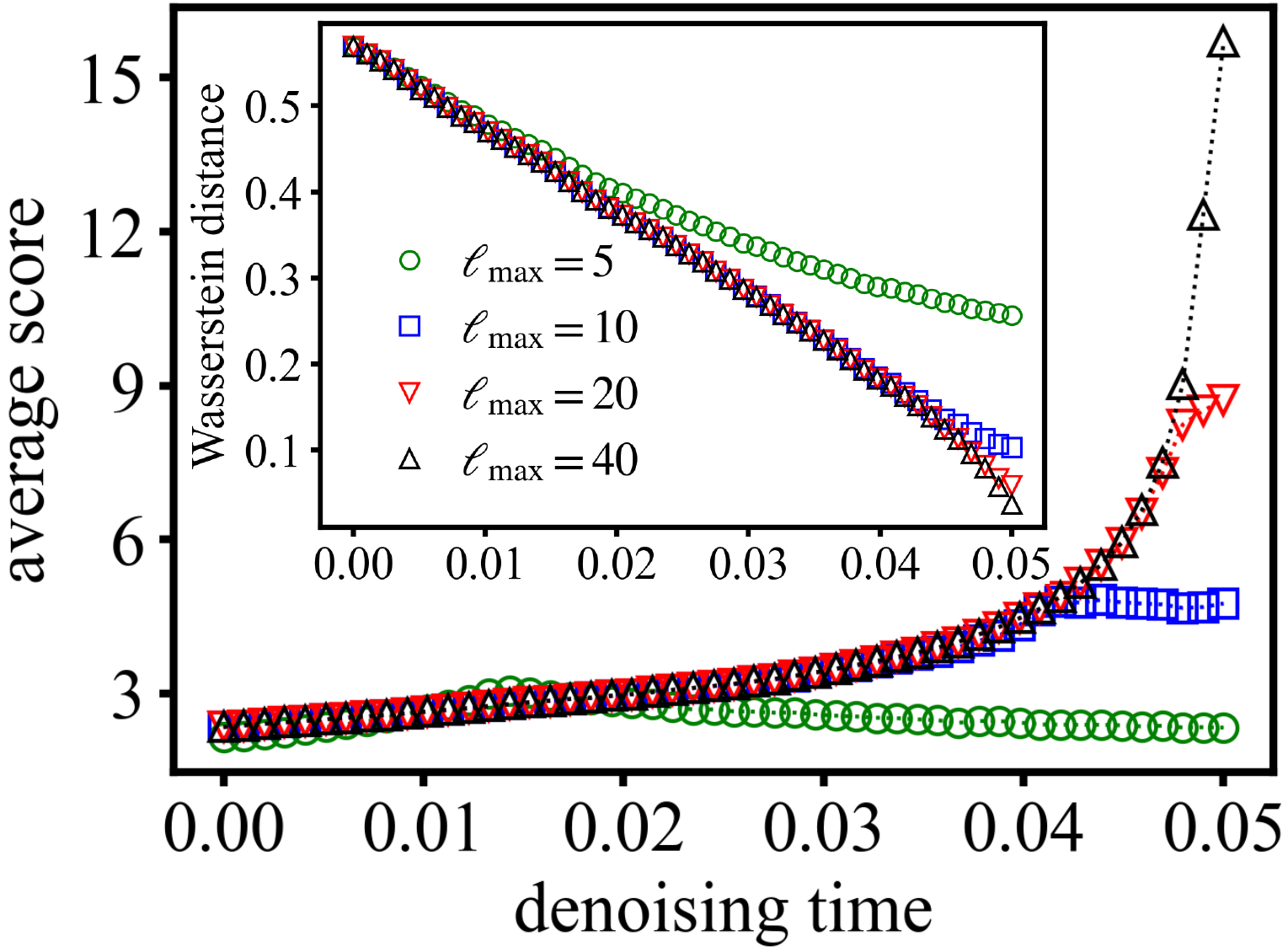}
  \caption{
    Average norm of the reverse Hamiltonian in \er{eq:example_reverse_Hamiltonian}
    as a function of denoising time in the two-level example, for several variational cut-offs $\ell_{\rm max}$. Inset: Wasserstein distance of the generated distribution of $\tilde{\psi}$ to the target distribution, as a function of denoising time. 
    \label{fig:Wasserstein_distance}
    }
\end{figure}

\smallskip
\noindent
{\em Level 2: Training with full signal but unknown initial states.} We now consider the case where we have the full measurement signal of $R$ sample forward trajectories, all starting from a copy of the same unknown initial state $\psi_{0}^{(n)} \sim \mu_0$, and we repeat this for a total of $N$ different independent initial states ($n=1,\ldots,N$) drawn from $\mu_0$. Even though each $\psi_{0}^{(n)}$ is unknown, the measurement records allow us to estimate it.

We denote $Y_t^{(n,r)} = (Y_{t,m}^{(n,r)})_{m=1}^M$ the measurement signals at time $t$ for sample trajectory $r$ started from the $n$-th initial state, and $I_t^{(n,r)}$ its corresponding current (so that $dY_t = I_t\,dt$). Under the model, $I_t^{(n,r)}$ admits the decomposition
\begin{equation}
  \label{eq:In}
  I_{t,m}^{(n,r)} 
  = 
  \Tr{L_m \, \psi_{t}^{(n,r)} + \psi_{t}^{(n,r)} L_m^\dagger} 
  + \xi_t^m
  \, ,
\end{equation}
where $\xi_t^m$ denotes independent white noise processes 
satisfying $\langle \xi_t^m \, \xi_{t'}^{m'} \rangle = 
\delta_{mm'}\delta(t-t').$

We define the observable space as
\begin{align}
\nonumber
\O
&=
\mathrm{span}
\Big\{
(\L^{*})^{a_1} \circ (\V_{m_1}^*)^{b_1} \circ
\cdots \circ
(\L^{*})^{a_k} \circ (\V_{m_k}^*)^{b_k}(\1)
:
\\
\nonumber
&
\qquad
m_i \in \{1,\ldots,M\},
\;
a_i,b_i\in\N_0,
\;
i=1,\ldots,k,
\;
k\in\N_0
\Big\}
\end{align}
where $\V_{m}(\cdot) = L_m (\cdot) + (\cdot) L_m^\dagger$ and $*$ indicates super-operator adjoint. The model is said to be {\em observable} if $\O$  coincides with $\B(\H)$. In such a case, the initial state can be identified from the measurement records, leveraging the correlation functions derived in \cite{guilmin2023correlation},
\begin{flalign}
  \label{eq:EI}
  \E[r]{I_{t_1,m_1}^{(n,r)} \cdots I_{t_k,m_k}^{(n,r)}}
  =
  & & 
  \\
  & &
  \mathllap{
    \Tr*{
    \psi_{0}^{(n)} \,
    e^{t_1 \L^*}
    \circ \V_{m_1}^*
    \circ \cdots
    \circ
    e^{(t_k-t_{k-1}) \L^*}
    \circ \V_{m_k}^*(\1)
    }
  }
  \nonumber
  \, ,
\end{flalign}
to estimate $\psi_{0}^{(n)}$ with statistical accuracy improving as the number of repetitions $R$ increases. 
By propagating the estimation of
$\psi_{0}^{(n)}$ with the measurement records $Y_t^{(n,r)},$ we reconstruct the whole forward trajectories $\psi^{(n,r)}_{t}$ and can optimise using Level 1 above. Note that under the observability condition above, purification holds, and hence so does filter stability \cite{amini2021on-asymptotic,bompais2026the-rate}, making the reconstruction robust to initial-state uncertainty.

\bigskip
\noindent
{\bf \em Example.} As an illustration of our general framework we consider the example 
of a qubit, $\H = \C^2$, where the noising dynamics is given by 
\ers{eq:SSE}{eq:Dissipator} with $H=0$, three coupling 
operators $L_{1,2,3} = \sigma_{x,y,z}$, and three independent Wiener processes $W^{x,y,z}_t.$ We take $\mu_0$ to be the uniform distribution 
over a great circle of the Bloch sphere with normal vector 
$\hat{n} = (1, 0, \sqrt{3})/2$.

The denoising dynamics \eqref{eq:reverse_SSE} in this case reads
\begin{align}
  d\tilde\psi_t 
  &= 
  \left(
    -i[{H}_t^\fb(\tilde\psi_t), \tilde\psi_t] 
    + \mathcal{D}(\tilde\psi_t)
  \right)dt 
  + \sum_{m=x,y,z}
    \mathcal{K}_m(\tilde\psi_t)\,d\tilde{W}_t^m
  \, ,
  \label{eq:example_reverse_SSE}
\end{align}
with $\mathcal{K}_m(\psi) = \sigma_m\psi + \psi\sigma_m - 2\operatorname{tr}(\sigma_m\psi)\,\psi$.
The explicit form of the reverse-time Hamiltonian is obtained by specialising \ers{eq:feedback_hamiltonian}{eq:partial_divergence} to this example. With $H=0$ we have $-2\mathcal{D}(\psi) = 4(2\psi - \1)$ 
and $\operatorname{div} D(\psi) = 4(\1 - 2\psi)$, and these two 
terms exactly cancel in $X_t(\psi)$. Note that this cancellation is specific to this example and does not occur in general. The only surviving 
contribution therefore comes from the score term, giving
\begin{equation}
  \label{eq:example_reverse_Hamiltonian}
  {H}_t^\fb(\psi) 
  = 
  2F(T-t,\,z_\psi)\;i\big[\hat{n}\cdot\vec{\sigma},\,\psi\big],
\end{equation}
where $z_\psi = \operatorname{tr}\!\big(\psi\,(\hat{n}\cdot\vec{\sigma})\big)$ is the projection of the Bloch vector onto $\hat{n}$, and $F(t, z) = \partial_z \log p(t, z)$, with $p(t,z)$ the probability density of $z_\psi$ under $\mu_t$.
The Hamiltonian \eqref{eq:example_reverse_Hamiltonian} is the direct quantum analogue of 
the score-based drift in classical reverse diffusions 
\cite{song2021score-based}: it is state-dependent, vanishes on the target ensemble $z_{\psi}=0,$ and its strength is modulated by 
the log-density gradient $F(t,z) = \partial_z \log p(t,z)$.\\

Figure~\ref{fig:bloch_spheres} illustrates this dynamics for the qubit example. 
In panel~(a), the forward noising dynamics progressively 
spreads the initial distribution $\mu_0$ over the great circle 
towards the invariant measure $\mu_{\mathrm{inv}}$, which is 
the uniform distribution over the Bloch sphere. In panel~(b), 
the reverse-time denoising dynamics reconstructs $\mu_0$ from 
samples of $\mu_{\mathrm{inv}}$, demonstrating that the 
reverse process correctly transports noise back to the 
target distribution.\\

Figure~\ref{fig:Wasserstein_distance} quantifies the convergence of the reverse-time 
denoising dynamics for the qubit example. The main panel 
shows the mean Frobenius norm of the reverse-time Hamiltonian,
$\mathbb{E}\big[\|{H}_t^\fb(\tilde\psi_t)\|_2\big]$,
as a function of backward time $t$, for several values of 
the truncation order $\ell_{\max}$ used in the score approximation.
The inset shows the Wasserstein distance 
$W_1(\tilde \mu_{t}, \mu_0)$ between the 
distribution of the backward state $\tilde\psi_t$ and the 
target distribution $\mu_0$, confirming convergence to $\mu_0$ 
as $t \to T$.
For details, see \cite{see-supplemental}.

\bigskip
\noindent
{\bf \em Outlook.} We have shown that every quantum diffusion associated with a continuously monitored quantum system admits a reverse-time evolution, which takes the form of a feedback-controlled quantum diffusion. This result can be leveraged to construct a sampling procedure for a target distribution of quantum states, as a direct generalisation of classical generative diffusion models.

A key feature of our framework is that randomness arises solely from the measurement process. The stochasticity is therefore intrinsic to the continuous observation of the system and does not rely on any additional external noise. Since quantum states cannot be directly accessed,  
the measurement process also plays an important additional role: it allows us to prepare a known  initial state for the denoising process, and to perform tomography of the sample states of the target distribution, when they are unknown. 

A natural extension of our framework will be to incorporate quantum model reduction techniques \cite{grigoletto2025quantum} to approximate high-dimensional dynamics by an effective lower-dimensional description. This could make the approach scalable to larger systems while preserving
the essential features of the stochastic evolution. Other interesting issues relate to studying in our quantum diffusive setup questions about ``memorisation'' of the training data, including both what occurs in time when running denoising dynamics on a trained model \cite{biroli2024dynamical}, or the conditions required for models trained with distinct sets of examples from the target to become equivalent in their generalising ability \cite{kadkhodaie2024generalization,maillard2026memorisation}.

\medskip

\begin{acknowledgments}
This work was supported by the Engineering and Physical Sciences Research Council, grants no. EP/T022140/1 and EP/V031201/1.
\end{acknowledgments}

\bibliography{bibliography-29052026.bib}

\clearpage
\setcounter{section}{0}
\onecolumngrid

\begin{center}
\textbf{\large Supplementary Material for}\\
\vspace{0.5em}
\textbf{Generating quantum ensembles via reverse-time quantum diffusions}
\end{center}

\section{Generator of the forward evolution}

The generator and adjoint generator of the quantum diffusion \eqref{eq:SSE} are the following (cf. \cite{carollo2021large}). For any scalar smooth test function $f$ on Hermitian matrices, the generator $\W$ acts as
\begin{equation}
(\W f)(\psi)
=
\L(\psi)\cdot \nabla f(\psi)
+
\frac12
\sum_{ij,hk}
D_{ij,hk}(\psi)\,
\frac{\partial^2 f}{\partial \psi_{ij}\partial \psi_{hk}}(\psi),
\end{equation}
where the diffusion tensor is given by
\begin{equation}
D_{ij,hk}(\psi)
=
\sum_{m=1}^M
\big(\K_m(\psi)\big)_{ij}
\big(\K_m(\psi)\big)_{hk}.
\end{equation}

The adjoint generator $\W^\dagger$ acts on probability densities $P(\psi)$ on Hermitian matrices as
\begin{equation}
(\mathcal W^\dagger P)(\psi)
=
-\,\nabla \cdot \big( \mathcal L(\psi)\, P(\psi) \big)
+
\frac12
\sum_{ij,hk}
\frac{\partial^2}{\partial \psi_{ij}\,\partial \psi_{hk}}
\big(
D_{ij,hk}(\psi)\,P(\psi)
\big).
\end{equation}

When $P$ is singular, derivatives are understood in the distributional sense. This is the case for the probability distributions associated with the quantum diffusion, which are supported on the manifold of pure states.

\section{Kolmogorov equations}

Let
$
p(s,x;t,y)
$
denote the transition kernel associated with a Markov process $(X_t)_{t\geq0}$, that is,
\begin{equation}
\mathbb P(X_t\in dy \mid X_s=x)
=
p(s,x;t,y)\,dy,
\qquad s<t.
\end{equation}

Under suitable regularity assumptions, the transition kernel satisfies the forward Kolmogorov equation
 (see, e.g., \cite{ikeda2014stochastic})
\begin{equation}
\partial_t p(s,x;t,y)
=
\mathcal W_t^{\dagger(y)}
\,p(s,x;t,y),
\end{equation}
where $\mathcal W_t^{\dagger(y)}$ denotes the formal adjoint of the (possibly time-dependent) generator $\mathcal W_t,$ acting on the variable $y$.

The transition kernel also satisfies the backward Kolmogorov equation
\begin{equation}
\partial_s p(s,x;t,y)
=
-
\mathcal W_s^{(x)}
\,p(s,x;t,y),
\end{equation}
where $\mathcal W_s^{(x)}$ denotes the generator $\mathcal W_s$ acting on the variable $x$.

\section{Deriving the reverse-time evolution}

\subsection{Derivation of the Bayes-type expression}

Let $(X_t)_{t\ge 0}$ be a Markov process with transition kernel
$p(s,x;t,y)$ and marginal distributions $\mu_t$. Let $\tilde p(s,x;t,y)$ be the transition kernel of the reversed dynamics,
i.e. of the process $(\tilde X_t)_{t \ge 0}$ defined by
\begin{equation}
\tilde X_t = X_{T-t}.
\end{equation}

By definition,
\begin{equation}
\tilde p(s,x;t,y)\,dy
=
\mathbb P(\tilde X_t \in dy \mid \tilde X_s = x).
\end{equation}

Using the definition of the time-reversed process,
\begin{equation}
\tilde p(s,x;t,y)\,dy
=
\mathbb P(X_{T-t} \in dy \mid X_{T-s} = x).
\end{equation}

By Bayes' formula,
\begin{equation}
\tilde p(s,x;t,y)\,dy
=
\frac{
\mathbb P(X_{T-t} \in dy \cap X_{T-s} \in dx)
}{
\mathbb P(X_{T-s} \in dx)
}.
\end{equation}

Since $\mathbb P(X_{T-s} \in dx) = \mu_{T-s}(x)\,dx$, we obtain
\begin{equation}
\tilde p(s,x;t,y)\,dy
=
\frac{
\mathbb P(X_{T-t} \in dy \cap X_{T-s} \in dx)
}{
\mu_{T-s}(x)\,dx
}.
\end{equation}

We now rewrite the joint probability in the correct time order:
\begin{equation}
\mathbb P(X_{T-t} \in dy \cap X_{T-s} \in dx)
=
\mathbb P(X_{T-s} \in dx \mid X_{T-t} \in dy)
\,
\mathbb P(X_{T-t} \in dy).
\end{equation}

By the Markov property,
\begin{equation}
\mathbb P(X_{T-s} \in dx \mid X_{T-t} = y)
=
p(T-t,y;T-s,x)\,dx,
\end{equation}
and
\begin{equation}
\mathbb P(X_{T-t} \in dy)
=
\mu_{T-t}(y)\,dy.
\end{equation}

Therefore,
\begin{equation}
\tilde p(s,x;t,y)\,dy
=
\frac{
p(T-t,y;T-s,x)\,\mu_{T-t}(y)\,dy\,dx
}{
\mu_{T-s}(x)\,dx
}.
\end{equation}

Cancelling $dx$ and identifying the densities with respect to $dy$, we obtain
\begin{equation}
\tilde p(s,x;t,y)
=
\frac{
p(T-t,y;T-s,x)\,\mu_{T-t}(y)
}{
\mu_{T-s}(x)
}.
\end{equation}

\subsection{Identification of the generator via Kolmogorov equations}

For pure states $\psi$ and $\phi$, let
$
p(s,\psi;t,\phi)
$
denote the transition kernel of the quantum diffusion \eqref{eq:SSE}, and let
$
\mu_t
$
denote the corresponding density at time $t$.
The reversed transition kernel is given by
\begin{equation}
\tilde p(s,\psi;t,\phi)
=
\frac{
p(T-t,\phi;T-s,\psi)\,\mu_{T-t}(\phi)
}{
\mu_{T-s}(\psi)
}.
\end{equation}
To identify the generator of the reversed process, we differentiate with respect to the first time argument.

\begin{align}
\frac{\partial}{\partial s}\tilde p(s,\psi;t,\phi)
&=
\frac{\partial}{\partial s}
\frac{
p(T-t,\phi;T-s,\psi)\,\mu_{T-t}(\phi)
}{
\mu_{T-s}(\psi)
} 
\end{align}

\medskip

The derivative is computed explicitly using the quotient rule.\\

\paragraph{Step 1. Quotient rule.}

Since $\mu_{T-t}(\phi)$ does not depend on $s$, we obtain
\begin{align}
\frac{\partial}{\partial s}\tilde p(s,\psi;t,\phi)
&=
\frac{\mu_{T-t}(\phi)}{\mu_{T-s}(\psi)}
\frac{\partial}{\partial s}
p(T-t,\phi;T-s,\psi)
\\
&\quad
-
\frac{
p(T-t,\phi;T-s,\psi)\,\mu_{T-t}(\phi)
}{
\mu_{T-s}(\psi)^2
}
\frac{\partial}{\partial s}
\mu_{T-s}(\psi).
\label{eq:dmuds}
\end{align}

\paragraph{Step 2. First term.}

Using the chain rule and the forward Kolmogorov equation for $p$:
\begin{align}
\frac{\partial}{\partial s}
p(T-t,\phi;T-s,\psi)
&=- \frac{\partial}{\partial u} p(T-t,\phi;u,\psi)\Big|_{u=T-s}\\
&=
\sum_{i,j}
\frac{\partial}{\partial \psi_{ij}}
\left(
\mathcal L(\psi)_{ij}
\,p(T-t,\phi;T-s,\psi)
\right)
\\
&\quad
-
\frac12
\sum_{i,j,h,k}
\frac{\partial^2}{\partial \psi_{ij}\partial \psi_{hk}}
\left(
D_{ij,hk}(\psi)\,
p(T-t,\phi;T-s,\psi)
\right).
\end{align}

Multiplying by $\frac{\mu_{T-t}(\phi)}{\mu_{T-s}(\psi)}$, we obtain
\begin{align}
\frac{\mu_{T-t}(\phi)}{\mu_{T-s}(\psi)}
\frac{\partial}{\partial s}
p(T-t,\phi;T-s,\psi)
&=
\sum_{i,j}
\frac{\mu_{T-t}(\phi)}{\mu_{T-s}(\psi)}
\frac{\partial}{\partial \psi_{ij}}
\left(
\mathcal L(\psi)_{ij}
\,p(T-t,\phi;T-s,\psi)
\right)
\\
&\quad
-
\frac12
\sum_{i,j,h,k}
\frac{\mu_{T-t}(\phi)}{\mu_{T-s}(\psi)}
\frac{\partial^2}{\partial \psi_{ij}\partial \psi_{hk}}
\left(
D_{ij,hk}(\psi)\,
p(T-t,\phi;T-s,\psi)
\right).
\end{align}

We now use that
$$
\frac{\mu_{T-t}(\phi)}{\mu_{T-s}(\psi)}
p(T-t,\phi;T-s,\psi)
=
\tilde p(s,\psi;t,\phi),
$$
and apply the product rule.

\medskip

\noindent First-order term:
\begin{align}
\frac{\mu_{T-t}(\phi)}{\mu_{T-s}(\psi)}
\frac{\partial}{\partial \psi_{ij}}
\left(
\mathcal L(\psi)_{ij}
\,p(T-t,\phi;T-s,\psi)
\right)
&=
\frac{\partial}{\partial \psi_{ij}}
\left(
\mathcal L(\psi)_{ij}\,
\tilde p(s,\psi;t,\phi)
\right)
\\
&\quad
+
\mathcal L(\psi)_{ij}\,
\tilde p(s,\psi;t,\phi)\,
\frac{\partial}{\partial \psi_{ij}}
\log \mu_{T-s}(\psi).
\end{align}

\medskip

\noindent Second-order term:
\begin{align}
\frac{\mu_{T-t}(\phi)}{\mu_{T-s}(\psi)}
\frac{\partial^2}{\partial \psi_{ij}\partial \psi_{hk}}
\left(
D_{ij,hk}(\psi)\,
p(T-t,\phi;T-s,\psi)
\right)
&=
\frac{\partial^2}{\partial \psi_{ij}\partial \psi_{hk}}
\left(
D_{ij,hk}(\psi)\,
\tilde p(s,\psi;t,\phi)
\right)
\\
&\quad
+
\frac{\partial}{\partial \psi_{ij}}
\left(
D_{ij,hk}(\psi)\,
\tilde p(s,\psi;t,\phi)\,
\frac{\partial}{\partial \psi_{hk}}
\log \mu_{T-s}(\psi)
\right)
\\
&\quad
+
\frac{\partial}{\partial \psi_{hk}}
\left(
D_{ij,hk}(\psi)\,
\tilde p(s,\psi;t,\phi)\,
\frac{\partial}{\partial \psi_{ij}}
\log \mu_{T-s}(\psi)
\right)
\\
&\quad
+
D_{ij,hk}(\psi)\,
\tilde p(s,\psi;t,\phi)
\\
&\quad\quad
\times
\left[
\frac{\partial}{\partial \psi_{ij}}\log \mu_{T-s}(\psi)\,
\frac{\partial}{\partial \psi_{hk}}\log \mu_{T-s}(\psi)
-
\frac{\partial^2}{\partial \psi_{ij}\partial \psi_{hk}}
\log \mu_{T-s}(\psi)
\right].
\end{align}

\paragraph{Step 3. Second term (derivative of $\mu_{T-s}$) in equation \eqref{eq:dmuds}.}

We have
\begin{align}
-
\frac{
p(T-t,\phi;T-s,\psi)\,\mu_{T-t}(\phi)
}{
\mu_{T-s}(\psi)^2
}
\frac{\partial}{\partial s}
\mu_{T-s}(\psi)
&=
-
\tilde p(s,\psi;t,\phi)
\frac{1}{\mu_{T-s}(\psi)}
\frac{\partial}{\partial s}
\mu_{T-s}(\psi).
\end{align}
Using the forward Kolmogorov equation, with the change of time variable,
\begin{align}
\frac{\partial}{\partial s}\mu_{T-s}(\psi)
&=
\sum_{i,j}
\frac{\partial}{\partial \psi_{ij}}
\left(
\mathcal L(\psi)_{ij}
\mu_{T-s}(\psi)
\right)
-
\frac12
\sum_{i,j,h,k}
\frac{\partial^2}{\partial \psi_{ij}\partial \psi_{hk}}
\left(
D_{ij,hk}(\psi)
\mu_{T-s}(\psi)
\right).
\end{align}
Therefore
\begin{align}
-
\tilde p(s,\psi;t,\phi)
\frac{1}{\mu_{T-s}(\psi)}
\frac{\partial}{\partial s}
\mu_{T-s}(\psi)
&=
-\tilde p(s,\psi;t,\phi)
\sum_{i,j}
\frac{1}{\mu_{T-s}(\psi)}
\frac{\partial}{\partial \psi_{ij}}
\left(
\mathcal L(\psi)_{ij}
\mu_{T-s}(\psi)
\right)
\\
&\quad
+
\frac12
\tilde p(s,\psi;t,\phi)
\sum_{i,j,h,k}
\frac{1}{\mu_{T-s}(\psi)}
\frac{\partial^2}{\partial \psi_{ij}\partial \psi_{hk}}
\left(
D_{ij,hk}(\psi)
\mu_{T-s}(\psi)
\right).
\end{align}
By the product rule,
\begin{align}
\frac{1}{\mu_{T-s}(\psi)}
\frac{\partial}{\partial \psi_{ij}}
\left(
\mathcal L(\psi)_{ij}
\mu_{T-s}(\psi)
\right)
&=
\frac{\partial \mathcal L(\psi)_{ij}}{\partial \psi_{ij}}
+
\mathcal L(\psi)_{ij}
\frac{\partial}{\partial \psi_{ij}}
\log\mu_{T-s}(\psi),
\end{align}
and
\begin{align}
\frac{1}{\mu_{T-s}(\psi)}
\frac{\partial^2}{\partial \psi_{ij}\partial \psi_{hk}}
\left(
D_{ij,hk}(\psi)
\mu_{T-s}(\psi)
\right)
&=
\frac{\partial^2 D_{ij,hk}(\psi)}
{\partial \psi_{ij}\partial \psi_{hk}}
+
\frac{\partial D_{ij,hk}(\psi)}{\partial \psi_{ij}}
\frac{\partial}{\partial \psi_{hk}}
\log\mu_{T-s}(\psi)
\\
&\quad
+
\frac{\partial D_{ij,hk}(\psi)}{\partial \psi_{hk}}
\frac{\partial}{\partial \psi_{ij}}
\log\mu_{T-s}(\psi)
\\
&\quad
+
D_{ij,hk}(\psi)
\frac{\partial^2}{\partial \psi_{ij}\partial \psi_{hk}}
\log\mu_{T-s}(\psi)
\\
&\quad
+
D_{ij,hk}(\psi)
\frac{\partial}{\partial \psi_{ij}}
\log\mu_{T-s}(\psi)
\frac{\partial}{\partial \psi_{hk}}
\log\mu_{T-s}(\psi).
\end{align}
Hence
\begin{align}
-
\tilde p(s,\psi;t,\phi)
\frac{1}{\mu_{T-s}(\psi)}
\frac{\partial}{\partial s}
\mu_{T-s}(\psi)
&=
-\tilde p(s,\psi;t,\phi)
\sum_{i,j}
\frac{\partial \mathcal L(\psi)_{ij}}{\partial \psi_{ij}}
\\
&\quad
-
\tilde p(s,\psi;t,\phi)
\sum_{i,j}
\mathcal L(\psi)_{ij}
\frac{\partial}{\partial \psi_{ij}}
\log\mu_{T-s}(\psi)
\\
&\quad
+
\frac12
\tilde p(s,\psi;t,\phi)
\sum_{i,j,h,k}
\frac{\partial^2 D_{ij,hk}(\psi)}
{\partial \psi_{ij}\partial \psi_{hk}}
\\
&\quad
+
\tilde p(s,\psi;t,\phi)
\sum_{i,j,h,k}
\frac{\partial D_{ij,hk}(\psi)}{\partial \psi_{ij}}
\frac{\partial}{\partial \psi_{hk}}
\log\mu_{T-s}(\psi)
\\
&\quad
+
\frac12
\tilde p(s,\psi;t,\phi)
\sum_{i,j,h,k}
D_{ij,hk}(\psi)
\frac{\partial^2}{\partial \psi_{ij}\partial \psi_{hk}}
\log\mu_{T-s}(\psi)
\\
&\quad
+
\frac12
\tilde p(s,\psi;t,\phi)
\sum_{i,j,h,k}
D_{ij,hk}(\psi)
\frac{\partial}{\partial \psi_{ij}}
\log\mu_{T-s}(\psi)
\frac{\partial}{\partial \psi_{hk}}
\log\mu_{T-s}(\psi),
\end{align}
where we used the symmetry $D_{ij,hk}=D_{hk,ij}$ to combine the two first-order terms involving derivatives of $D$.\\

\paragraph{Step 4. Final cancellation and identification.}

Combining this with the contribution obtained in Step 2, we get
\begin{align}
\frac{\partial}{\partial s}\tilde p(s,\psi;t,\phi)
&=
\sum_{i,j}
\mathcal L(\psi)_{ij}
\frac{\partial}{\partial \psi_{ij}}
\tilde p(s,\psi;t,\phi)
\\
&\quad
-
\sum_{i,j,h,k}
\frac{\partial D_{ij,hk}(\psi)}{\partial \psi_{ij}}
\frac{\partial}{\partial \psi_{hk}}
\tilde p(s,\psi;t,\phi)
\\
&\quad
-
\sum_{i,j,h,k}
D_{ij,hk}(\psi)
\frac{\partial}{\partial \psi_{ij}}
\log\mu_{T-s}(\psi)
\frac{\partial}{\partial \psi_{hk}}
\tilde p(s,\psi;t,\phi)
\\
&\quad
-
\frac12
\sum_{i,j,h,k}
D_{ij,hk}(\psi)
\frac{\partial^2}{\partial \psi_{ij}\partial \psi_{hk}}
\tilde p(s,\psi;t,\phi).
\end{align}
Equivalently,
\begin{align}
\frac{\partial}{\partial s}\tilde p(s,\psi;t,\phi)
&=
- \sum_{i,j}
\tilde{\mathcal L}_s(\psi)_{ij}
\frac{\partial}{\partial \psi_{ij}}
\tilde p(s,\psi;t,\phi)
-
\frac12
\sum_{i,j,h,k}
D_{ij,hk}(\psi)
\frac{\partial^2}{\partial \psi_{ij}\partial \psi_{hk}}
\tilde p(s,\psi;t,\phi),
\end{align}
where
\begin{align}
\tilde{\mathcal L}_s(\psi)_{ij}
&=
- \mathcal L(\psi)_{ij}
+
\sum_{h,k}
\frac{\partial D_{ij,hk}(\psi)}{\partial \psi_{hk}}
+
\sum_{h,k}
D_{ij,hk}(\psi)
\frac{\partial}{\partial \psi_{hk}}
\log\mu_{T-s}(\psi).
\end{align}

This is exactly the backward Kolmogorov equation:
\begin{equation}
\frac{\partial}{\partial s}\tilde p(s,\psi;t,\phi)
=
-  \widetilde \W^{(\psi)}_{s}
\big(
\tilde p(s,\psi;t,\phi)
\big)
\end{equation}
where $\widetilde \W_s$ is given by
\begin{equation}
\widetilde{\mathcal W}_s
=
\sum_{i,j}
\tilde{\mathcal L}_s(\psi)_{ij}
\frac{\partial}{\partial \psi_{ij}}
+
\frac12
\sum_{i,j,h,k}
D_{ij,hk}(\psi)
\frac{\partial^2}{\partial \psi_{ij}\partial \psi_{hk}}.
\end{equation}

This identifies the generator of the evolution of $\tilde\psi_t$. The corresponding stochastic differential equation reads
\begin{equation}\label{eq:raw_bwd}
d \tilde \psi_t
=
\Big(
- \L(\tilde \psi_t)
+
\operatorname{div} D(\tilde \psi_t)
+
D(\tilde \psi_t)\,
\nabla \log \mu_{T-t}(\tilde \psi_t)
\Big)\,dt
+
\sum_m \K_m(\tilde \psi_t)\,d \tilde W_t^m,
\end{equation}
where $\tilde W_t^m$ are independent Wiener processes and 
\begin{equation}
(\operatorname{div} D)_{ij}(\psi)
=
\sum_{h,k}
\frac{\partial}{\partial \psi_{hk}}
\big(
D_{ij,hk}(\psi)
\big).
\end{equation}

\subsection{Hamiltonian representation of the backward drift}

\subsubsection{Characterisation of commutator directions}

Let $\psi = \ketbra{\psi}$ be a rank-one projector, so that $\psi^2 = \psi$, and let $X(\psi) \in \B(\H)$ be Hermitian. Then the following are equivalent:
\begin{enumerate}
    \item There exists a Hermitian operator $H(\psi)$ such that
    \begin{equation}
    X(\psi) = -i[H(\psi),\psi].
    \end{equation}
    
    \item $X(\psi)$ satisfies
    \begin{equation}
    \psi X(\psi)\psi = 0,
    \qquad
    (\1 - \psi) X(\psi)(\1 - \psi) = 0.
    \end{equation}
\end{enumerate}

\paragraph{Proof.}

$(1) \Rightarrow (2)$.

Assume $X(\psi) = -i[H(\psi),\psi]$. Then
\begin{align}
\psi X(\psi) \psi
&= -i\, \psi\big(H(\psi)\psi - \psi H(\psi)\big)\psi \\
&= -i\big(\psi H(\psi)\psi - \psi H(\psi)\psi\big) = 0,
\end{align}
where we used $\psi^2 = \psi$. Moreover,
\begin{align}
(\1 - \psi) X(\psi) (\1 - \psi)
&= -i\, (\1 - \psi)\big(H(\psi)\psi - \psi H(\psi)\big)(\1 - \psi) \\
&= 0,
\end{align}
since $(\1 - \psi)\psi = \psi(\1 - \psi) = 0$.

\medskip

$(2) \Rightarrow (1)$.

Assume
\begin{equation}
\psi X(\psi)\psi = 0,
\qquad
(\1 - \psi) X(\psi)(\1 - \psi) = 0.
\end{equation}

We first decompose $X(\psi)$ using $\1 = \psi + (\1 - \psi)$:
\begin{align}
X(\psi)
&= (\psi + (\1 - \psi))\, X(\psi)\, (\psi + (\1 - \psi)) \\
&= \psi X(\psi)\psi
+ \psi X(\psi)(\1 - \psi)
+ (\1 - \psi) X(\psi)\psi
+ (\1 - \psi) X(\psi)(\1 - \psi).
\end{align}
Using the assumptions, we obtain
\begin{equation}
X(\psi)
= \psi X(\psi)(\1 - \psi)
+ (\1 - \psi) X(\psi)\psi.
\end{equation}

Define
\begin{equation}
H(\psi) := i[X(\psi),\psi].
\end{equation}
Since $X(\psi)$ and $\psi$ are Hermitian,
\begin{equation}
[X(\psi),\psi]^\dagger
= (X(\psi)\psi-\psi X(\psi))^\dagger
= \psi X(\psi)-X(\psi)\psi
= -[X(\psi),\psi].
\end{equation}
Hence $[X(\psi),\psi]$ is anti-Hermitian, and therefore $H(\psi)$ is Hermitian.

Moreover,
\begin{align}
-i[H(\psi),\psi]
&= -i\big(i[X(\psi),\psi]\psi - i\psi[X(\psi),\psi]\big) \\
&= [X(\psi),\psi]\psi - \psi[X(\psi),\psi].
\end{align}
Expanding the commutators gives
\begin{align}
-i[H(\psi),\psi]
&= \big(X(\psi)\psi - \psi X(\psi)\big)\psi
 - \psi\big(X(\psi)\psi - \psi X(\psi)\big) \\
&= X(\psi)\psi - \psi X(\psi)\psi
 - \psi X(\psi)\psi + \psi X(\psi).
\end{align}
Using $\psi^2=\psi$, we obtain
\begin{align}
-i[H(\psi),\psi]
&= (\1-\psi)X(\psi)\psi
+ \psi X(\psi)(\1-\psi).
\end{align}
By the decomposition obtained above,
\begin{equation}
X(\psi)
=
(\1-\psi)X(\psi)\psi
+
\psi X(\psi)(\1-\psi),
\end{equation}
and therefore
\begin{equation}
-i[H(\psi),\psi] = X(\psi).
\end{equation}

\hfill $\square$

\subsubsection{Verification of the condition $\psi X(\psi)\psi = 0$}

The backward drift of \eqref{eq:raw_bwd} can be written in the form
\begin{equation}
- i [H, \psi] + \mathcal D(\psi) -i[-2H,\psi] + X(\psi),
\end{equation}
with
\begin{equation}
X(\psi)
=
- 2 \mathcal D(\psi)
+ \operatorname{div} D(\psi)
+ D(\psi)\nabla \log \mu_{T-t}(\psi).
\end{equation}
We now check that
\begin{equation}
\psi X(\psi)\psi = 0
\end{equation}
for every pure state $\psi$.

\medskip

Since $\psi$ is a pure state, it is a rank-one orthogonal projection, so that
\begin{equation}
\psi^2 = \psi,
\qquad
\tr{\psi}=1,
\qquad
\psi A \psi = \tr{A\psi}\,\psi
\quad
\text{for every operator } A.
\end{equation}

We analyse separately the three terms entering the definition of $X(\psi)$.

\paragraph{The score term.}

Write
\begin{equation}
D(\psi)\nabla \log \mu_{T-t}(\psi)
=
\sum_{m=1}^M \K_m(\psi)\, S_m(\psi,t),
\end{equation}
for suitable scalar functions \(S_m(\psi,t)\). Therefore
\begin{equation}
\psi\, D(\psi)\nabla \log \mu_{T-t}(\psi)\,\psi
=
\sum_{m=1}^M S_m(\psi,t)\, \psi \K_m(\psi)\psi.
\end{equation}
It is thus enough to show that
\begin{equation}
\psi \K_m(\psi)\psi = 0
\qquad
\text{for every } m.
\end{equation}

Fix $m$, and set
\begin{equation}
\alpha_m := \tr(L_m\psi),
\qquad
\bar\alpha_m = \tr(L_m^\dag\psi),
\qquad
\lambda_m := \tr((L_m+L_m^\dag)\psi)=\alpha_m+\bar\alpha_m.
\end{equation}
Then
\begin{align}
\psi \K_m(\psi)\psi
&=
\psi L_m\psi\psi + \psi\psi L_m^\dag\psi - \lambda_m \psi^3 \\
&=
\psi L_m\psi + \psi L_m^\dag\psi - \lambda_m \psi \\
&=
\tr(L_m\psi)\psi + \tr(L_m^\dag\psi)\psi - (\alpha_m+\bar\alpha_m)\psi \\
&= 0.
\end{align}
Hence
\begin{equation}
\psi\, D(\psi)\nabla \log \mu_{T-t}(\psi)\,\psi = 0.
\end{equation}

\paragraph{The dissipative term.}

Recall that
\begin{equation}
\mathcal D(\psi)
=
\sum_{m=1}^M \mathcal D_m(\psi),
\qquad
\mathcal D_m(\psi)
:=
L_m\psi L_m^\dag - \frac12\{L_m^\dag L_m,\psi\}.
\end{equation}
We compute $\psi \mathcal D_m(\psi)\psi$ for fixed $m$.

Using again that $\psi A\psi = \tr(A\psi)\psi$, we obtain
\begin{align}
\psi L_m\psi L_m^\dag \psi
&=
(\psi L_m\psi)(\psi L_m^\dag \psi) \\
&=
\tr(L_m\psi)\tr(L_m^\dag\psi)\,\psi \\
&=
|\alpha_m|^2\,\psi.
\end{align}
Moreover,
\begin{align}
\psi \{L_m^\dag L_m,\psi\}\psi
&=
\psi (L_m^\dag L_m \psi + \psi L_m^\dag L_m)\psi \\
&=
\psi L_m^\dag L_m \psi + \psi L_m^\dag L_m \psi \\
&=
2\,\tr(L_m^\dag L_m\psi)\,\psi.
\end{align}
If we set
\begin{equation}
\beta_m := \tr(L_m^\dag L_m\psi),
\end{equation}
this gives
\begin{equation}
\psi \mathcal D_m(\psi)\psi
=
\big(|\alpha_m|^2-\beta_m\big)\psi.
\end{equation}
Therefore
\begin{equation}
\psi \mathcal D(\psi)\psi
=
\sum_{m=1}^M \big(|\alpha_m|^2-\beta_m\big)\psi.
\end{equation}

\paragraph{The divergence term.}

Recall that
\begin{equation}
D_{ij,hk}(\psi)
=
\sum_{m=1}^M \K_m(\psi)_{ij}\K_m(\psi)_{hk}.
\end{equation}
Hence
\begin{equation}
\operatorname{div} D(\psi)
=
\sum_{m=1}^M \operatorname{div}\big(\K_m\otimes \K_m\big)(\psi).
\end{equation}
For a vector field $K$, one has component-wise
\begin{equation}
\big(\operatorname{div}(K\otimes K)\big)_{hk}
=
\sum_{i,j}\frac{\partial}{\partial \psi_{ij}}\big(K_{ij}K_{hk}\big)
=
\Big(\sum_{i,j}\frac{\partial K_{ij}}{\partial \psi_{ij}}\Big)K_{hk}
+
\sum_{i,j}K_{ij}\frac{\partial K_{hk}}{\partial \psi_{ij}}.
\end{equation}
In other words,
\begin{equation}
\operatorname{div}(K\otimes K)
=
(\operatorname{div}K)\,K + dK[K].
\end{equation}
Applying this with $K=\K_m$, we obtain
\begin{equation}
\operatorname{div} D(\psi)
=
\sum_{m=1}^M
\Big(
(\operatorname{div}\K_m(\psi))\,\K_m(\psi)
+
d\K_m(\psi)[\K_m(\psi)]
\Big).
\end{equation}
Since $\operatorname{div}\K_m(\psi)$ is a scalar and $\psi \K_m(\psi)\psi=0$, it follows that
\begin{equation}
\psi\, (\operatorname{div}\K_m(\psi))\,\K_m(\psi)\,\psi = 0.
\end{equation}
Thus
\begin{equation}
\psi\, \operatorname{div} D(\psi)\,\psi
=
\sum_{m=1}^M
\psi\, d\K_m(\psi)[\K_m(\psi)]\,\psi.
\end{equation}

We now compute $d\K_m(\psi)[\delta]$ for an arbitrary variation $\delta$. Since
\begin{equation}
\K_m(\psi)
=
L_m\psi + \psi L_m^\dag - \lambda_m(\psi)\psi,
\qquad
\lambda_m(\psi)=\tr((L_m+L_m^\dag)\psi),
\end{equation}
we get
\begin{equation}
d\K_m(\psi)[\delta]
=
L_m\delta + \delta L_m^\dag
- \tr((L_m+L_m^\dag)\delta)\,\psi
- \lambda_m(\psi)\,\delta.
\end{equation}
We now evaluate this at $\delta=\K_m(\psi)$ and sandwich by $\psi$:
\begin{align}
\psi\, d\K_m(\psi)[\K_m(\psi)]\,\psi
&=
\psi L_m\K_m(\psi)\psi
+
\psi \K_m(\psi)L_m^\dag\psi \\
&\quad
-
\tr((L_m+L_m^\dag)\K_m(\psi))\,\psi
-
\lambda_m(\psi)\,\psi\K_m(\psi)\psi.
\end{align}
The last term vanishes because $\psi\K_m(\psi)\psi=0$, so
\begin{equation}
\psi\, d\K_m(\psi)[\K_m(\psi)]\,\psi
=
\psi L_m\K_m(\psi)\psi
+
\psi \K_m(\psi)L_m^\dag\psi
-
\tr((L_m+L_m^\dag)\K_m(\psi))\,\psi.
\label{mention_1}
\end{equation}

We now simplify each term. First,
\begin{align}
\K_m(\psi)\psi
&=
L_m\psi\psi + \psi L_m^\dag\psi - \lambda_m \psi^2 \\
&=
L_m\psi + \bar\alpha_m \psi - (\alpha_m+\bar\alpha_m)\psi \\
&=
L_m\psi - \alpha_m\psi.
\end{align}
Therefore,
\begin{align}
\psi L_m\K_m(\psi)\psi
&=
\psi L_m(L_m\psi-\alpha_m\psi) \\
&=
\psi L_m^2\psi - \alpha_m \psi L_m\psi \\
&=
\tr(L_m^2\psi)\psi - \alpha_m^2 \psi.
\end{align}
Similarly,
\begin{align}
\psi\K_m(\psi)
&=
\psi L_m\psi + \psi^2 L_m^\dag - \lambda_m \psi^2 \\
&=
\alpha_m \psi + \psi L_m^\dag - (\alpha_m+\bar\alpha_m)\psi \\
&=
\psi L_m^\dag - \bar\alpha_m \psi,
\end{align}
hence
\begin{align}
\psi \K_m(\psi)L_m^\dag\psi
&=
(\psi L_m^\dag - \bar\alpha_m\psi)L_m^\dag\psi \\
&=
\psi (L_m^\dag)^2\psi - \bar\alpha_m \psi L_m^\dag\psi \\
&=
\tr((L_m^\dag)^2\psi)\psi - \bar\alpha_m^2 \psi.
\end{align}

It remains to compute $\tr((L_m+L_m^\dag)\K_m(\psi))$. Using the definition of $\K_m(\psi)$,
\begin{align}
\tr((L_m+L_m^\dag)\K_m(\psi))
&=
\tr((L_m+L_m^\dag)L_m\psi)
+
\tr((L_m+L_m^\dag)\psi L_m^\dag)
-
\lambda_m\,\tr((L_m+L_m^\dag)\psi).
\end{align}
Since
\begin{equation}
\tr((L_m+L_m^\dag)\psi)=\lambda_m,
\end{equation}
and using cyclicity of the trace, this becomes
\begin{align}
\tr((L_m+L_m^\dag)\K_m(\psi))
&=
\tr(L_m^2\psi)
+
\tr(L_m^\dag L_m\psi)
+
\tr(L_m^\dag L_m\psi)
+
\tr((L_m^\dag)^2\psi)
-
\lambda_m^2 \\
&=
\tr(L_m^2\psi)
+
2\beta_m
+
\tr((L_m^\dag)^2\psi)
-
\lambda_m^2.
\end{align}

Substituting the above expressions into \eqref{mention_1}, we obtain
\begin{align}
\psi\, d\K_m(\psi)[\K_m(\psi)]\,\psi
&=
\Big(
\tr(L_m^2\psi)-\alpha_m^2
+
\tr((L_m^\dag)^2\psi)-\bar\alpha_m^2 \\
&\qquad
-
\tr(L_m^2\psi)
-
2\beta_m
-
\tr((L_m^\dag)^2\psi)
+
\lambda_m^2
\Big)\psi \\
&=
\big(
-\alpha_m^2 - \bar\alpha_m^2 - 2\beta_m + \lambda_m^2
\big)\psi.
\end{align}
Now, since $\lambda_m=\alpha_m+\bar\alpha_m$, we have
\begin{equation}
\lambda_m^2-\alpha_m^2-\bar\alpha_m^2 = 2|\alpha_m|^2,
\end{equation}
and therefore
\begin{equation}
\psi\, d\K_m(\psi)[\K_m(\psi)]\,\psi
=
2\big(|\alpha_m|^2-\beta_m\big)\psi.
\end{equation}
Summing over $m$, we conclude that
\begin{equation}
\psi\, \operatorname{div} D(\psi)\,\psi
=
2\sum_{m=1}^M \big(|\alpha_m|^2-\beta_m\big)\psi
=
2\,\psi \mathcal D(\psi)\psi.
\end{equation}

\paragraph{Conclusion.}

Putting everything together, we obtain
\begin{align}
\psi X(\psi)\psi
&=
-2\,\psi\mathcal D(\psi)\psi
+\psi\,\operatorname{div} D(\psi)\,\psi
+\psi\, D(\psi)\nabla \log \mu_{T-t}(\psi)\,\psi \\
&=
- 2\,\psi\mathcal D(\psi)\psi + 2\,\psi\mathcal D(\psi)\psi + 0 \\
&= 0.
\end{align}
Hence the first condition is satisfied:
\begin{equation}
\boxed{\psi X(\psi)\psi = 0.}
\end{equation}

\subsubsection{Verification of the condition $(\1-\psi) X(\psi) (\1-\psi) = 0$}

We now verify the second condition
\begin{equation}
(\1-\psi) X(\psi) (\1-\psi) = 0.
\end{equation}

Recall that
\begin{equation}
X(\psi)
=
- 2 \mathcal D(\psi)
+ \operatorname{div} D(\psi)
+ D(\psi)\nabla \log \mu_{T-t}(\psi),
\end{equation}
where
\begin{equation}
\mathcal D(\psi)
=
\sum_{m=1}^M
\left(
L_m \psi L_m^\dag
- \frac12 \{L_m^\dag L_m,\psi\}
\right),
\end{equation}
and
\begin{equation}
D(\psi)\nabla \log \mu_{T-t}(\psi)
=
\sum_{m=1}^M \K_m(\psi)\, S_m(\psi,t),
\end{equation}
with
\begin{equation}
\K_m(\psi)
=
L_m \psi + \psi L_m^\dag
-
\tr((L_m+L_m^\dag)\psi)\,\psi.
\end{equation}

For simplicity, let us set
\begin{equation}
P := \psi,
\qquad
Q := \1-\psi.
\end{equation}
Since $\psi$ is a pure state, $P$ is a rank-one orthogonal projection, and
\begin{equation}
P^2=P,
\qquad
Q^2=Q,
\qquad
PQ=QP=0.
\end{equation}

We analyse separately the three terms entering the definition of $X(\psi)$.\\

\paragraph{The score term.}

Recall that
\begin{equation}
D(P)\nabla \log \mu_{T-t}(P)
=
\sum_{m=1}^M \K_m(P)\, S_m(P,t),
\end{equation}
where each $S_m(P,t)$ is a scalar. Thus
\begin{equation}
Q\, D(P)\nabla \log \mu_{T-t}(P)\, Q
=
\sum_{m=1}^M S_m(P,t)\, Q \K_m(P) Q.
\end{equation}
It is therefore enough to prove that
\begin{equation}
Q \K_m(P) Q = 0
\qquad
\text{for every } m.
\end{equation}

Fix $m$, and set
\begin{equation}
\lambda_m := \tr((L_m+L_m^\dag)P).
\end{equation}
Then
\begin{align}
Q \K_m(P) Q
&=
Q\big(L_mP + PL_m^\dag - \lambda_m P\big)Q \\
&=
QL_mPQ + QPL_m^\dag Q - \lambda_m QPQ.
\end{align}
Again, since $PQ=QP=0$, all three terms are zero:
\begin{equation}
QL_mPQ = 0,
\qquad
QPL_m^\dag Q = 0,
\qquad
QPQ=0.
\end{equation}
Hence
\begin{equation}
Q \K_m(P) Q = 0,
\end{equation}
and therefore
\begin{equation}
Q\, D(P)\nabla \log \mu_{T-t}(P)\, Q = 0.
\end{equation}

\paragraph{The dissipative term.}

We write
\begin{equation}
\mathcal D(P)
=
\sum_{m=1}^M \mathcal D_m(P),
\qquad
\mathcal D_m(P)
=
L_m P L_m^\dag - \frac12\{L_m^\dag L_m,P\}.
\end{equation}
We compute $Q \mathcal D_m(P) Q$ for fixed $m$:
\begin{align}
Q \mathcal D_m(P) Q
&=
Q L_m P L_m^\dag Q
-
\frac12 Q(L_m^\dag L_m P + P L_m^\dag L_m)Q.
\end{align}
Now,
\begin{equation}
Q L_m^\dag L_m P Q = Q L_m^\dag L_m (PQ) = 0,
\end{equation}
and
\begin{equation}
Q P L_m^\dag L_m Q = (QP)L_m^\dag L_m Q = 0.
\end{equation}
Thus the anticommutator term vanishes after sandwiching by $Q$, and we obtain
\begin{equation}
Q \mathcal D_m(P) Q = Q L_m P L_m^\dag Q.
\end{equation}
Summing over $m$ gives
\begin{equation}
Q \mathcal D(P) Q
=
\sum_{m=1}^M Q L_m P L_m^\dag Q.
\end{equation}

\paragraph{The divergence term.}

Recall that
\begin{equation}
D_{ij,hk}(P)
=
\sum_{m=1}^M \K_m(P)_{ij}\K_m(P)_{hk},
\end{equation}
hence
\begin{equation}
\operatorname{div} D(P)
=
\sum_{m=1}^M \operatorname{div}(\K_m\otimes \K_m)(P).
\end{equation}
For a vector field $K$, one has componentwise
\begin{equation}
\operatorname{div}(K\otimes K)
=
(\operatorname{div}K)\,K + dK[K].
\end{equation}
Therefore,
\begin{equation}
\operatorname{div} D(P)
=
\sum_{m=1}^M
\Big(
(\operatorname{div}\K_m(P))\,\K_m(P)
+
d\K_m(P)[\K_m(P)]
\Big).
\end{equation}
Since $\operatorname{div}\K_m(P)$ is a scalar and $Q\K_m(P)Q=0$, we immediately obtain
\begin{equation}
Q\, (\operatorname{div}\K_m(P))\,\K_m(P)\, Q = 0.
\end{equation}
Thus
\begin{equation}
Q\, \operatorname{div} D(P)\, Q
=
\sum_{m=1}^M
Q\, d\K_m(P)[\K_m(P)]\, Q.
\end{equation}

It remains to compute $d\K_m(P)[\K_m(P)]$.

\medskip

Let $\delta$ be an arbitrary variation. Since
\begin{equation}
\K_m(\psi)
=
L_m \psi + \psi L_m^\dag - \lambda_m(\psi)\,\psi,
\qquad
\lambda_m(\psi)=\tr((L_m+L_m^\dag)\psi),
\end{equation}
we have
\begin{equation}
d\K_m(P)[\delta]
=
L_m \delta + \delta L_m^\dag
-
\tr((L_m+L_m^\dag)\delta)\,P
-
\lambda_m(P)\,\delta.
\end{equation}
Evaluating at $\delta=\K_m(P)$, we find
\begin{equation}
d\K_m(P)[\K_m(P)]
=
L_m \K_m(P)
+
\K_m(P)L_m^\dag
-
\tr((L_m+L_m^\dag)\K_m(P))\,P
-
\lambda_m(P)\,\K_m(P).
\end{equation}
Now sandwich by $Q$:
\begin{align}
Q\, d\K_m(P)[\K_m(P)]\, Q
&=
Q L_m \K_m(P) Q
+
Q \K_m(P) L_m^\dag Q \\
&\quad
-
\tr((L_m+L_m^\dag)\K_m(P))\, Q P Q
-
\lambda_m(P)\, Q \K_m(P) Q.
\end{align}
The last two terms vanish because $QPQ=0$ and $Q\K_m(P)Q=0$. Therefore
\begin{equation}
Q\, d\K_m(P)[\K_m(P)]\, Q
=
Q L_m \K_m(P) Q
+
Q \K_m(P) L_m^\dag Q.
\label{mention_2}
\end{equation}

We now simplify the two remaining terms.

First,
\begin{align}
\K_m(P)Q
&=
\big(L_mP + PL_m^\dag - \lambda_m P\big)Q \\
&=
L_m P Q + P L_m^\dag Q - \lambda_m P Q.
\end{align}
Since $PQ=0$, this reduces to
\begin{equation}
\K_m(P)Q = P L_m^\dag Q.
\end{equation}
Hence
\begin{equation}
Q L_m \K_m(P) Q
=
Q L_m P L_m^\dag Q.
\end{equation}

Similarly,
\begin{align}
Q\K_m(P)
&=
Q\big(L_mP + PL_m^\dag - \lambda_m P\big) \\
&=
Q L_m P + Q P L_m^\dag - \lambda_m QP.
\end{align}
Since $QP=0$, this reduces to
\begin{equation}
Q\K_m(P)=Q L_m P.
\end{equation}
Therefore
\begin{equation}
Q \K_m(P) L_m^\dag Q
=
Q L_m P L_m^\dag Q.
\end{equation}

Substituting these two identities into \eqref{mention_2}, we obtain
\begin{equation}
Q\, d\K_m(P)[\K_m(P)]\, Q
=
2\, Q L_m P L_m^\dag Q.
\end{equation}
Summing over $m$ gives
\begin{equation}
Q\, \operatorname{div} D(P)\, Q
=
2 \sum_{m=1}^M Q L_m P L_m^\dag Q.
\end{equation}
Comparing with the expression obtained above for $Q\mathcal D(P)Q$, we conclude that
\begin{equation}
Q\, \operatorname{div} D(P)\, Q
=
2\, Q \mathcal D(P) Q.
\end{equation}

\paragraph{Conclusion.}

Putting everything together, we find
\begin{align}
Q X(P) Q
&=
- 2 Q \mathcal D(P) Q
+ Q\, \operatorname{div} D(P)\, Q
+ Q\, D(P)\nabla \log \mu_{T-t}(P)\, Q \\
&=
- 2 Q \mathcal D(P) Q + 2 Q \mathcal D(P) Q + 0 \\
&= 0.
\end{align}
Since $Q=\1-\psi$ and $P=\psi$, this proves that
\begin{equation}
\boxed{
(\1-\psi)\, X(\psi)\, (\1-\psi)=0.
}
\end{equation}

\section{Example}

\subsection{Derivation of $-2\mathcal{D}(\psi)$ and $\operatorname{div}D(\psi)$ in the example}

In the example we have $H=0$ and $L_m = \sigma_m$ for $m \in \{x,y,z\}$,
with $\sigma_m^\dagger = \sigma_m$ and $\sigma_m^2 = {\1} $.
We use throughout $\psi^2 = \psi$ and $\tr(\psi)=1$, together with 
the following two identities for $2\times 2$ matrices, both consequences 
of $\sum_m (\sigma_m)_{ij}(\sigma_m)_{kl} = 2\delta_{il}\delta_{jk} - \delta_{ij}\delta_{kl}$:
\begin{align}
    \sum_{m} \sigma_m A \sigma_m &= 2\tr(A)\,{\1}  - A,
    \label{eq:pauli-sandwich}\\
    \sum_m \tr(\sigma_m A)\,\sigma_m &= 2A - \tr(A)\,{\1} ,
    \label{eq:pauli-expansion}
\end{align}
obtained by contracting with $A_{lk}$ and summing over $k,l$
in each index slot respectively.

\paragraph{Term $-2\mathcal{D}(\psi)$.}
Since $\sigma_m^\dagger = \sigma_m$ and $\sigma_m^2 = {\1} $,
\begin{equation}
    \mathcal{D}(\psi) 
    = \sum_m \left(\sigma_m\psi\sigma_m - \psi\right) 
    = \sum_m \sigma_m\psi\sigma_m - 3\psi.
\end{equation}
Applying \eqref{eq:pauli-sandwich} with $A=\psi$ and using $\tr(\psi)=1$:
\begin{equation}
    \mathcal{D}(\psi) = 2{\1}  - \psi - 3\psi = 2{\1}  - 4\psi,
\end{equation}
so
\begin{equation}
    -2\mathcal{D}(\psi) = 4(2\psi - {\1} ).
    \label{eq:minus2D}
\end{equation}

\paragraph{Term $\operatorname{div}D(\psi)$.}
By the product rule applied to 
$D_{ij,hk} = \sum_m [\mathcal{K}_m]_{ij}[\mathcal{K}_m]_{hk}$:
\begin{equation}
(\operatorname{div}D)_{ij} 
= \sum_m \left(
[\mathcal{K}_m]_{ij}\, \alpha_m
+ \beta_{m,ij}
\right),
\label{eq:divD-split}
\end{equation}
where $\alpha_m \coloneqq \sum_{h,k}\partial_{\psi_{hk}}[\mathcal{K}_m]_{hk}$
and $\beta_{m,ij} \coloneqq \sum_{h,k}[\mathcal{K}_m]_{hk}\,\partial_{\psi_{hk}}[\mathcal{K}_m]_{ij}$.
From $\mathcal{K}_m(\psi) = \sigma_m\psi + \psi\sigma_m - 2\tr(\sigma_m\psi)\psi$:
\begin{equation}
\frac{\partial [\mathcal{K}_m]_{ij}}{\partial \psi_{hk}} 
= (\sigma_m)_{ih}\delta_{jk} + \delta_{ih}(\sigma_m)_{kj} 
- 2(\sigma_m)_{kh}\psi_{ij} - 2\tr(\sigma_m\psi)\,\delta_{ih}\delta_{jk}.
\label{eq:dKdpsi}
\end{equation}

\noindent\textit{Term $\alpha_m$.}
Setting $i=h$, $j=k$ in \eqref{eq:dKdpsi} and summing,
and using $\tr(\sigma_m)=0$, $\tr({\1} )=2$:
\begin{equation}
\alpha_m = -10\tr(\sigma_m\psi).
\end{equation}
The contribution $\sum_m \alpha_m \mathcal{K}_m = -10\sum_m \tr(\sigma_m\psi)\,\mathcal{K}_m$ 
vanishes. Indeed, applying \eqref{eq:pauli-expansion} with $A=\psi$
and using $\psi^2=\psi$, $\tr(\psi)=1$:
\begin{align}
\sum_m \tr(\sigma_m\psi)\,\mathcal{K}_m
&= (2\psi-{\1} )\psi + \psi(2\psi-{\1} ) 
   - 2\tr\!\big((2\psi-{\1} )\psi\big)\psi \nonumber\\
&= 4\psi^2 - 2\psi - 2(2\tr\psi^2 - \tr\psi)\psi \nonumber\\
&= 4\psi - 2\psi - 2(2-1)\psi = 0.
\end{align}

\noindent\textit{Term $\beta_{m,ij}$.}
Contracting \eqref{eq:dKdpsi} with $[\mathcal{K}_m]_{hk}$ and summing over $h,k$:
\begin{equation}
\beta_{m,ij} = (\sigma_m\mathcal{K}_m)_{ij} + (\mathcal{K}_m\sigma_m)_{ij}
- 2\tr(\mathcal{K}_m\sigma_m)\,\psi_{ij} 
- 2\tr(\sigma_m\psi)\,[\mathcal{K}_m]_{ij}.
\end{equation}
Summing over $m$:
\begin{itemize}
\item \textit{$\sum_m(\sigma_m\mathcal{K}_m + \mathcal{K}_m\sigma_m)$}: 
expanding using $\mathcal{K}_m = \sigma_m\psi+\psi\sigma_m - 2\tr(\sigma_m\psi)\psi$,
applying \eqref{eq:pauli-sandwich} and $\psi^2=\psi$:
\begin{equation}
\sum_m(\sigma_m\mathcal{K}_m + \mathcal{K}_m\sigma_m) 
= 6\psi + 2(2{\1} -\psi) - 2\cdot 2\psi = 4{\1} ,
\end{equation}
where we used $\sum_m\sigma_m\psi\sigma_m = 2{\1} -\psi$ 
and $\sum_m\tr(\sigma_m\psi)(\sigma_m\psi+\psi\sigma_m) 
= (2\psi-{\1} )\psi + \psi(2\psi-{\1} ) = 4\psi-2\psi = 2\psi$.

\item \textit{$-2\sum_m\tr(\mathcal{K}_m\sigma_m)\psi$}: 
using $\tr(\mathcal{K}_m\sigma_m) 
= \tr((\sigma_m\psi+\psi\sigma_m)\sigma_m) - 2\tr(\sigma_m\psi)\tr(\sigma_m\psi)
= 2\tr(\psi) - 2\tr(\sigma_m\psi)^2 = 2 - 2\tr(\sigma_m\psi)^2$:
\begin{equation}
-2\sum_m\tr(\mathcal{K}_m\sigma_m)\psi 
= -2\sum_m\big(2 - 2\tr(\sigma_m\psi)^2\big)\psi
= \big(-12 + 4\sum_m\tr(\sigma_m\psi)^2\big)\psi.
\end{equation}
Now $\sum_m\tr(\sigma_m\psi)^2 = \tr\!\left(\psi\sum_m\tr(\sigma_m\psi)\sigma_m\right) 
= \tr(\psi(2\psi-{\1} )) = 2\tr\psi^2 - \tr\psi = 1$,
so this term equals $-8\psi$.

\item \textit{$-2\sum_m\tr(\sigma_m\psi)\mathcal{K}_m = 0$} as shown above.
\end{itemize}

\noindent Combining all contributions:
\begin{equation}
    \operatorname{div}D(\psi) = 4{\1}  - 8\psi = 4({\1}  - 2\psi).
    \label{eq:divD-result}
\end{equation}

\paragraph{Cancellation.}
From \eqref{eq:minus2D} and \eqref{eq:divD-result}:
\begin{equation}
    -2\mathcal{D}(\psi) + \operatorname{div}D(\psi) 
    = 4(2\psi-{\1} ) + 4({\1} -2\psi) = 0.
\end{equation}

\subsection{Derivation of the reverse-time Hamiltonian in the example}

We choose $\mu_0$ as the uniform distribution over the great circle of the
Bloch sphere with normal vector $\hat{n}$. We set $N := \hat{n}\cdot\vec{\sigma}$
for brevity. We derive the explicit form of the reverse-time Hamiltonian
$H^\fb_t(\psi) = i[X_t(\psi),\psi]$, where
\begin{equation}
X_t(\psi)
=
-2\mathcal{D}(\psi)
+\operatorname{div}D(\psi)
+D(\psi)\nabla_\psi\log\mu_{T-t}(\psi).
\label{eq:X-example}
\end{equation}

\paragraph{Step 1: Cancellation of the first two terms.}

As shown in the previous section,
\begin{equation}
-2\mathcal{D}(\psi) + \operatorname{div}D(\psi)
= 4(2\psi-{\1} ) + 4({\1} -2\psi) = 0,
\end{equation}
so that
\begin{equation}
X_t(\psi) = D(\psi)\nabla_\psi\log\mu_{T-t}(\psi).
\label{eq:X-score-only}
\end{equation}

\paragraph{Step 2: Evaluation of the score term.}

By symmetry of the dynamics and of $\mu_0$ under rotations about $\hat{n}$,
the distribution $\mu_t$ depends on $\psi$ only through the scalar
\begin{equation}
z_\psi := \tr\!\big(\psi\,N\big).
\end{equation}
Writing $\mu_t(\psi) = \frac{1}{2 \pi} p(t,z_\psi)$ for a probability density $p$, we have
\begin{equation}
\frac{\partial \log\mu_{T-t}(\psi)}{\partial \psi_{hk}}
=
F(T-t,z_\psi)\,N_{kh},
\qquad
F(t,z) := \partial_z \log p(t,z).
\end{equation}
Therefore, using the definition of the diffusion tensor,
\begin{equation}
[D(\psi)\nabla_\psi\log\mu_{T-t}(\psi)]_{ij}
=
F\sum_m [\mathcal{K}_m(\psi)]_{ij}\,\tr\!\big(\mathcal{K}_m(\psi)\,N\big).
\label{eq:score-contraction}
\end{equation}
We compute $\tr(\mathcal{K}_m(\psi)N)$. Using
$\mathcal{K}_m(\psi) = \sigma_m\psi + \psi\sigma_m - 2\tr(\sigma_m\psi)\psi$
and cyclicity of the trace:
\begin{align}
\tr\!\big((\sigma_m\psi+\psi\sigma_m)N\big)
&=
\tr(\sigma_m\psi N) + \tr(\psi\sigma_m N)
\nonumber\\
&=
\tr\!\big(\psi(N\sigma_m+\sigma_m N)\big).
\end{align}
Using the Pauli anticommutation relation
$\{\sigma_m, N\} = 2n_m{\1} $,
which follows from $\{\sigma_m,\sigma_n\}=2\delta_{mn}{\1} $, we obtain
\begin{equation}
\tr\!\big((\sigma_m\psi+\psi\sigma_m)N\big)
=
2n_m\tr(\psi) = 2n_m.
\end{equation}
Therefore
\begin{equation}
\tr\!\big(\mathcal{K}_m(\psi)\,N\big)
=
2n_m - 2\tr(\sigma_m\psi)\,z_\psi.
\label{eq:trKN}
\end{equation}
Substituting \eqref{eq:trKN} into \eqref{eq:score-contraction}:
\begin{equation}
X_t(\psi)
=
2F(T-t,z_\psi)
\sum_m
\Big(n_m - \tr(\sigma_m\psi)\,z_\psi\Big)\,\mathcal{K}_m(\psi).
\end{equation}
The term $\sum_m\tr(\sigma_m\psi)\,\mathcal{K}_m(\psi)$ vanishes, as shown
in the previous section. Therefore
\begin{equation}
X_t(\psi)
=
2F(T-t,z_\psi)\sum_m n_m\,\mathcal{K}_m(\psi).
\label{eq:X-nm}
\end{equation}
Expanding using $\mathcal{K}_m(\psi) = \sigma_m\psi+\psi\sigma_m-2\tr(\sigma_m\psi)\psi$
and $\sum_m n_m\sigma_m = N$, $\sum_m n_m\tr(\sigma_m\psi) = \tr(N\psi) = z_\psi$:
\begin{align}
\sum_m n_m\,\mathcal{K}_m(\psi)
&=
N\psi + \psi N - 2z_\psi\psi
=
\{N - z_\psi{\1} ,\,\psi\},
\end{align}
so that
\begin{equation}
X_t(\psi)
=
2F(T-t,z_\psi)\,\{N-z_\psi{\1} ,\,\psi\}.
\label{eq:X-anticommutator}
\end{equation}

\paragraph{Step 3: Explicit form of $H^\fb_t(\psi)$.}

We compute $[X_t(\psi),\psi]$ using \eqref{eq:X-anticommutator},
$\psi^2=\psi$, and $\psi N\psi = z_\psi\psi$:
\begin{align}
[X_t(\psi),\psi]
&=
2F\big[N\psi+\psi N - 2z_\psi\psi,\,\psi\big]
\nonumber\\
&=
2F\big(N\psi^2 + \psi N\psi - \psi N\psi - \psi^2 N - 2z_\psi\psi^2 + 2z_\psi\psi^2\big)
\nonumber\\
&=
2F\big(N\psi + z_\psi\psi - z_\psi\psi - \psi N\big)
\nonumber\\
&=
2F\,[N,\psi].
\end{align}
Therefore the reverse-time Hamiltonian is
\begin{equation}
H^\fb_t(\psi)
=
2F(T-t,z_\psi)\,i\big[\hat{n}\cdot\vec{\sigma},\,\psi\big].
\label{eq:tH-correct}
\end{equation}

\subsection{Expression of the score function $F(t,z)$}
In the example considered here, the initial distribution $\mu_0$
is invariant under rotations around the axis $\hat n$.
Since the dynamics generated by the Pauli operators
$\sigma_x,\sigma_y,\sigma_z$ is isotropic, this symmetry is preserved
for all times. Consequently, the distribution $\mu_t$ depends on
$\psi$ only through the scalar
\begin{equation}
z_\psi = \tr(\psi N),
\qquad
N=\hat n\cdot\vec{\sigma},
\end{equation}
and can therefore be written as
\begin{equation}
\mu_t(\psi)= \frac{1}{2 \pi}p(t,z_\psi)
\end{equation}
for a one-dimensional probability density $p(t,z)$ on $[-1,1]$.
The quantity
\begin{equation}
F(t,z)=\partial_z\log p(t,z)
\end{equation}
appearing in the reverse-time Hamiltonian is thus the score function
associated with this one-dimensional density.

Applying It\^o's formula to $z_\psi = \tr(\psi N)$ along the forward SSE
and using \eqref{eq:trKN}, one finds that $z_t$ satisfies the
\emph{Jacobi diffusion}
\begin{equation}
    dz_t = -4z_t\,dt + 2\sqrt{1-z_t^2}\,dW_t
\end{equation}
on $[-1,1]$, with generator $\mathcal{G} = -4z\,\partial_z + 2(1-z^2)\partial_{zz}$.
The Legendre polynomials are eigenfunctions of $\mathcal{G}$, with
$\mathcal{G}P_\ell = -\lambda_\ell P_\ell$ and eigenvalues
$\lambda_\ell = 2\ell(\ell+1)$,
as one verifies directly from the Legendre equation
$(1-z^2)P_\ell'' - 2zP_\ell' = -\ell(\ell+1)P_\ell$.
The stationary measure of $z_t$ is the uniform distribution $dz/2$ on $[-1,1]$,
with respect to which the $P_\ell$ are orthogonal.

For the particular initial condition considered here, namely the
uniform distribution on the great circle orthogonal to $\hat n$,
every state on this circle satisfies $z_\psi = 0$, so
$p(0,z) = \delta(z)$.
The solution of the Fokker--Planck equation associated with
$\mathcal G$, with initial condition $p(0,z)=\delta(z)$,
admits the spectral expansion
\begin{equation}
p(t,z)
=
\frac12
\sum_{\ell=0}^{\infty}
(2\ell+1)
P_\ell(0)\,
P_\ell(z)\,
e^{-\lambda_\ell t},
\end{equation}
where the coefficients $(2\ell+1)P_\ell(0)$ follow from
$\int_{-1}^1 \delta(z) P_\ell(z)\,dz = P_\ell(0)$
and the orthogonality relation
$\int_{-1}^1 P_j P_\ell\,\frac{dz}{2} = \frac{1}{2\ell+1}\delta_{j\ell}$.
Note that $P_\ell(0)=0$ for odd $\ell$, so only even-degree terms contribute.
Differentiating term-by-term yields
\begin{equation}
F(t,z)
=
\frac{
\displaystyle
\sum_{\ell=0}^{\infty}
(2\ell+1)
P_\ell(0)\,
P_\ell'(z)\,
e^{-\lambda_\ell t}
}{
\displaystyle
\sum_{\ell=0}^{\infty}
(2\ell+1)
P_\ell(0)\,
P_\ell(z)\,
e^{-\lambda_\ell t}
}.
\label{eq:F-explicit}
\end{equation}
Substituting \eqref{eq:F-explicit} into
\eqref{eq:tH-correct} provides a fully explicit expression for the
reverse-time Hamiltonian.\\

In practice, the numerical simulations use the truncated approximation
\begin{equation}
F(t,z)
\approx
\frac{
\displaystyle
\sum_{\ell=0}^{\ell_{\max}}
(2\ell+1)
P_\ell(0)\,
P_\ell'(z)\,
e^{-\lambda_\ell t}
}{
\displaystyle
\sum_{\ell=0}^{\ell_{\max}}
(2\ell+1)
P_\ell(0)\,
P_\ell(z)\,
e^{-\lambda_\ell t}
},
\end{equation}
where $\ell_{\max}$ is a finite cutoff. This approximation is justified
for any $t>0$ by the exponential decay of the coefficients
$e^{-\lambda_\ell t}$, with $\lambda_\ell = 2\ell(\ell+1)$.

\end{document}